%% file: AliHFMpp5dot02.tex
\documentclass[ALICE,manyauthors]{cernphprep}
\usepackage{AliHFMpp5dot02}

\DeclareUnicodeCharacter{00A0}{}

\begin{document}

\begin{titlepage}
\PHyear{2019}
\PHnumber{078}        
\PHdate{19 April}  

\title{Production of muons from heavy-flavour hadron decays in pp collisions at $\sqrt{s}=5.02$~TeV}
\ShortTitle{HF decay $\mu$ production in pp collisions at $\sqrt s$ = 5.02 TeV}  

\Collaboration{ALICE Collaboration\thanks{See Appendix~\ref{app:collab} for the list of collaboration members}}
\ShortAuthor{ALICE Collaboration} 

\begin{abstract}
\input{body/c00Abstract.tex}
\end{abstract}
\end{titlepage}

\setcounter{page}{2}

\section{Introduction}\label{sec:intro}
\input{body/c01Introduction.tex}

\section{Experimental apparatus and data taking conditions}\label{sec:det}
\input{body/c02DetData.tex}

\section{Data analysis}\label{sec:strat}
\input{body/c03Analysis.tex}

\section{Results and comparison with model predictions}\label{sec:res}
\input{body/c04Results.tex}

\section{Conclusions}\label{sec:concl}
\input{body/c05Conclusion.tex}

\newpage
\newenvironment{acknowledgement}{\relax}{\relax}
\begin{acknowledgement}
\section*{Acknowledgements}
\input{fa_2019-03-23}
\end{acknowledgement}

\bibliographystyle{utphys} 
\bibliography{AliHFMpp5dot02}

\newpage
\appendix

\section{The ALICE Collaboration}
\label{app:collab}
\input{2019-03-23-Alice_Authorlist_2019-Mar-23}
\end{document}

%% file: body/c00Abstract.tex

Production cross sections of muons from semi-leptonic decays of charm and 
beauty hadrons were measured at forward rapidity ($2.5<y<4$) in 
proton--proton (pp) collisions at a centre-of-mass energy 
$\sqrt{s}=5.02$~TeV with the ALICE detector at the CERN LHC. 
The results were obtained in an extended transverse momentum interval, 
$2 < p_{\rm T} < 20$~GeV/$c$, 
and with an improved precision compared to previous measurements performed 
in the same rapidity interval at centre-of-mass energies $\sqrt{s}= 2.76$ and 
7~TeV.
The $\pT$- and $y$-differential production cross sections as well as 
the $\pT$-differential production cross section ratios between different 
centre-of-mass energies and different rapidity intervals 
are described, within experimental and theoretical 
uncertainties, by predictions based on perturbative QCD.

%% file: body/c01Introduction.tex

The measurement of heavy-flavour (charm and beauty) production cross sections 
in proton--proton (pp) collisions at the CERN LHC represents an important 
test of perturbative Quantum Chromodynamics (pQCD). Due to their large 
masses, heavy quarks are produced almost exclusively in initial hard partonic 
scatterings and consequently their production cross sections can be estimated 
in the framework of pQCD. The calculations are based on a factorisation 
approach and computed as a convolution of the hard parton scattering 
cross section, evaluated as a perturbative series of the coupling constant 
of the strong interaction, the parton distribution function (PDF) of the 
colliding protons and the fragmentation function of heavy quarks to 
heavy-flavour hadrons. 
Heavy-flavour production cross sections are predicted 
at next-to-leading order (NLO) using the fixed-order plus 
next-to-leading logarithms (FONLL) 
approach~\cite{Cacciari:1998it,Cacciari:2012ny} or 
the general-mass variable-flavour-number scheme 
(GM-VFNS)~\cite{Kniehl:2008eu,Kniehl:2011bk}. Calculations at leading order 
based on 
$k_{\rm T}$ factorisation~\cite{Maciula:2013oba} also exist. 
The forward rapidity range accessible by ALICE ($2.5 < y <4$) 
allows us to test pQCD predictions in a region of small Bjorken $x$ down to 
about $10^{-5}$ ($x$ being the longitudinal momentum 
fraction of initial-state partons, primarily gluons). 
In this region, the gluon distribution functions 
are affected by large uncertainties~\cite{Cacciari:2015fta}. 
The systematic uncertainties on the theoretical production cross sections 
are larger than the experimental ones and are dominated by the uncertainties 
on renormalisation and factorisation scales. Recent theoretical developments 
have shown that the 
ratios of the open heavy-flavour production cross sections 
between different beam energies and different rapidity intervals 
are promising observables which are 
expected to be sensitive to the gluon PDFs~\cite{Cacciari:2015fta}, since the 
uncertainties on scales become negligible with respect to the 
PDF uncertainties when calculating such ratios. 
The production cross sections of charm, beauty and heavy-flavour hadron 
decay leptons measured over a wide energy domain at the Tevatron, 
RHIC 
and LHC (see e.g.~\cite{Andronic:2015wma} and references therein and, 
\cite{Khachatryan:2016csy,Acharya:2018upq,Abelev:2012sca,Acharya:2017kfy,
Acharya:2017lwf,Acharya:2017jgo,Adam:2016ich,Aaij:2015bpa,Aaij:2016jht}) are 
described, within uncertainties, by these pQCD-based calculations  
at both forward and central rapidities in a large transverse momentum 
($p_{\rm T}$) range. Also the ratios of D-meson production 
cross sections between different rapidity 
intervals and centre-of mass energies recently measured by the ALICE and 
LHCb experiments~\cite{Acharya:2017jgo,Aaij:2015bpa,Aaij:2016jht} are 
described by pQCD-based predictions within uncertainties.

Furthermore, the measurement 
of heavy-flavour production cross sections 
in pp collisions provides the necessary baseline for the corresponding 
measurements in proton--nucleus and nucleus--nucleus collisions. 
These measurements allow us to study cold nuclear 
matter effects and effects related to the hot strongly-interacting 
medium, respectively.

This letter describes the $p_{\rm T}$- and $y$-differential measurements 
of the production 
cross sections of muons from the decay of charm and beauty 
hadrons in pp collisions at 
$\sqrt s$ = 5.02 TeV, with the ALICE detector at the LHC. 
These measurements are performed at forward rapidity, in the interval 
$2.5<y<4$. They are facilitated by an experimentally triggerable observable 
and relatively
large decay branching ratios (about 10\%), thus resulting in relatively large 
statistics allowing for differential measurements over a wide $p_{\rm T}$ 
interval. The present measurements cover the interval 
$2 <p_{\rm T} < 20$~GeV/$c$, where the 
beauty contribution is expected to dominate over the charm contribution in 
the high $p_{\rm T}$ region i.e. for 
$p_{\rm T} > 5$ GeV/$c$~\cite{Cacciari:2012ny}. 
They are complementary to 
those performed at the same centre-of-mass energy by the LHCb Collaboration 
for D-meson species in a kinematic region limited to 
hadron $p_{\rm T}<10$~GeV/$c$~\cite{Aaij:2016jht}. 
Moreover, the present results are obtained in a significantly extended 
$p_{\rm T}$ region and the total uncertainties are reduced by a factor 
larger than two, compared to previous published ALICE 
results for muons from heavy-flavour hadron 
decays~\cite{Abelev:2012qh,Abelev:2012pi}. 

The letter is structured as follows. Section~\ref{sec:det} 
describes the apparatus with an emphasis on the detectors used in the 
analysis and the data taking 
conditions. Section~\ref{sec:strat} addresses the analysis 
details. 
Section~\ref{sec:res} presents the 
results, namely the $\pT$- and $y$-differential cross sections of muons from 
heavy-flavour hadron decays as well as the ratio of the $\pT$-differential 
cross section between different centre-of-mass energies and rapidity intervals 
and their comparison 
with pQCD-based FONLL calculations. 
Finally, conclusions are drawn in Section~\ref{sec:concl}.

%% file: body/c02DetData.tex

The ALICE detector and its performance are described in detail 
in~\cite{Aamodt:2008zz,Abelev:2014ffa}.
This analysis is based on muons reconstructed in the muon spectrometer which 
covers the pseudo-rapidity interval $-4<\eta_{\rm lab}<-2.5$\footnote{The 
muon spectrometer covers a negative pseudo-rapidity range in the ALICE 
reference                 
frame. $\eta$ and $y$ variables are experimentally identical for muons 
in the acceptance of 
the muon spectrometer and in pp collisions the physics results 
are symmetric with respect to $\eta$ ($y$) = 0. They are presented 
as a function of $y$ with positive values.} in the laboratory 
frame.
The muon spectrometer consists of
i) a front absorber made of carbon, concrete and steel of 10 nuclear 
interaction 
lengths ($\lambda_{\rm I}$), located between the interaction point (IP) and 
the tracking system, that reduces the hadron yield and decreases the yield 
of muons from light-particle 
decays by limiting the free path of primary pions and kaons,
ii) a beam shield throughout its entire length,
iii) a dipole magnet with a field integral of 3 T$\cdot$m,
iv) five tracking stations, each composed of two planes of cathode pad 
chambers, v) two trigger stations,
each equipped with two planes of resistive plate chambers and vi) an iron 
wall of $7.2$~$\lambda_{\rm I}$ placed between the tracking and trigger 
systems, which absorbs secondary hadrons escaping from the front absorber 
as well as muons from light-hadron decays.
In addition, the following detectors are also employed in the analysis.
The Silicon Pixel Detector (SPD), which constitutes the two innermost layers 
of the Inner Tracking System, with pseudo-rapidity 
coverage $\vert\eta_{\rm lab}\vert<2$ and $\vert\eta_{\rm lab}\vert<1.4$ 
for the inner and outer layer, respectively, is used for reconstructing 
the position of the interaction vertex.
Two scintillator arrays (V0) placed on each side of the IP, with 
pseudo-rapidity coverage $2.8<\eta_{\rm lab}<5.1$ and 
$-3.7<\eta_{\rm lab}<-1.7$, are used for triggering purposes and to reject 
offline beam-induced background events. 
Finally, the two T0 arrays, made of quartz Cerenkov counters and placed on 
both sides 
of the IP, covering the acceptance $4.6<\eta_{\rm lab}<4.9$ and 
$-3.3<\eta_{\rm lab}<-3.0$, are employed to determine the luminosity.

The results presented in this letter are based on the pp data sample at a 
centre-of-mass energy $\sqrt s$ = 5.02 TeV recorded by ALICE during a short data taking period of five days in November $2015$.
This data sample consists of muon-triggered events requiring the coincidence 
of the minimum-bias (MB) trigger condition and at least one track segment 
in the muon trigger system with a $\pT$ above the threshold of 
the online trigger algorithm. The MB trigger is formed by a coincidence between signals in the two V0 arrays.
The samples of single muons were collected with the $\pT$ threshold of 
the online trigger algorithm set to provide 
a 50\% efficiency for muon tracks with either $\pT$~$\sim$~0.5~GeV/$c$ or 
$\pT$~$\sim$~4.2~GeV/$c$. 
In the following, the low- and high-$\pT$ trigger threshold samples are 
referred to as MSL and MSH, respectively. Beam-gas interactions are 
reduced at the offline level using the timing information of the V0 detector. 
The accepted events have at least 
one interaction vertex reconstructed from 
hits correlation in the two SPD layers.
The pile-up rate, defined as the probability for multiple interactions in a 
bunch crossing, was smaller than $2.5\%$ during the whole data taking period 
and taken into account in the luminosity determination. 
After the event selection described above, 
the integrated luminosities for the used data 
samples are ${\mathcal L}_{\rm int}$ = $53.7\pm 1.1$~nb$^{-1}$ and 
${\mathcal L}_{\rm int}$ = $104.4\pm 2.2~$nb$^{-1}$  for MSL- 
and MSH-triggered events, respectively. 
The calculation of the integrated luminosities 
and associated uncertainties 
is discussed in Section~\ref{sec:strat}.

%% file: body/c03Analysis.tex

\subsection{Selection of muon candidates}\label{subsec:selec}

Muon candidates are reconstructed using the algorithm described 
in~\cite{Aamodt:2011gj}. They are further selected for the analysis 
applying same offline criteria as those described in 
\cite{Abelev:2012pi,Abelev:2012qh}. 
The muon identification is performed by requiring that the reconstructed 
track in the tracking system matches a track segment in the trigger system 
satisfying the trigger condition. Muon candidates are required to be 
reconstructed in the pseudo-rapidity region $-4 < \eta_{\rm lab} < -2.5$ 
and to have a polar angle measured at the end of the absorber in 
the interval $170^\circ < \theta_{\rm abs} < 178^\circ$. 
The $\theta_{\rm abs}$ condition 
allows us to limit multiple scattering by rejecting tracks passing 
through the high-density part of the front absorber. The contamination of 
fake tracks coming from the 
association of uncorrelated clusters in the tracking chambers 
and beam-induced background tracks is further reduced by applying a 
selection on the distance of the track to the primary vertex measured in 
the transverse plane (DCA, distance of closest approach) weighted with 
its momentum ($p$). 
The maximum value is set to $6 \sigma_{p \cdot {\rm DCA}}$, 
where $\sigma_{p \cdot {\rm DCA}}$ is the resolution on this quantity. 
Finally, only muons with $p_{\rm T} > 2$~GeV/$c$ are analysed since 
according to Monte Carlo simulations~\cite{Abelev:2012pi}, the 
contribution of muons from the decay of secondary light hadrons produced 
inside the front absorber is expected to be small in this 
region. The statistics recorded by ALICE allows us to 
perform the measurement of the production of muons from 
heavy-flavour hadron decays up to $p_{\rm T}$ = 20 GeV/$c$ by combining 
MSL- and MSH-triggered events, which are used up to and above 
$p_{\rm T}$ = 7 GeV/$c$, respectively. 
In the selected interval 
$2 < p_{\rm T} < 20$~GeV/$c$, the main remaining background contributions 
consist of muons from the decay of light (charged) hadrons (mostly pions and 
kaons) produced at the IP and muons from 
W and Z/$\gamma^{*}$ decays, which dominate at low/intermediate $p_{\rm T}$ 
($p_{\rm T} < 6-7$~GeV/$c$) and high $\pT$ ($p_{\rm T} > 16-17$~GeV/$c$), 
respectively. Moreover, two additional background contributions, muons from secondary 
light (charged) hadron decays and muons from J/$\psi$ decays, are also taken 
into account in the analysis, although they are small 
compared to the two other background sources.

\subsection{Analysis procedure}\label{subsec:proc}

The differential production cross section of muons from 
heavy-flavour hadron decays in a given $\pT$ and $y$ interval is computed as:
\begin{linenomath}
\begin{equation}
{{{\rm d^2}\sigma^{\mu^\pm\leftarrow {\rm {HF}}}} \over 
{{\rm d}p_{\rm T} {\rm d}y }} = 
{{\rm d^2\sigma^{\mu^\pm}} \over {{\rm d}p_{\rm T} {\rm d}y }} 
- {{{\rm d^2}\sigma^{\mu^\pm\leftarrow {\rm \pi}}} \over 
{{\rm d}p_{\rm T}{\rm d}y }} 
- {{{\rm d}^2\sigma^{\mu^\pm\leftarrow {\rm K}}} \over 
{{\rm d}p_{\rm T}{\rm d}y }} 
- {{{\rm d^2}\sigma^{\mu^\pm\leftarrow {\rm sec. \pi/K}}} \over 
{{\rm d}p_{\rm T}{\rm d}y }} 
- {{{\rm d^2}\sigma^{\mu^\pm\leftarrow {\rm W/Z/\gamma^*}}} \over 
{{\rm d}p_{\rm T}{\rm d}y }} 
- {{{\rm d^2}\sigma^{\mu^\pm\leftarrow {\rm J/\psi}}} \over 
{{\rm d}p_{\rm T}{\rm d}y }},
\label{eq:CrossSecHF}
\end{equation}
\end{linenomath}
where ${\rm d^2}\sigma^{\mu^\pm} / {\rm d}p_{\rm T}{\rm d}y $ is the 
$\pT$- and $y$-differential production cross section of inclusive muons and, 
${\rm d^2}\sigma^{\mu^\pm\leftarrow {\rm \pi}} / {\rm d}p_{\rm T}{\rm d}y$, 
${\rm d^2}\sigma^{\mu^\pm\leftarrow {\rm K}} / {\rm d}p_{\rm T}{\rm d}y$, 
${\rm d^2}\sigma^{\mu^\pm\leftarrow {\rm sec. \pi/K}} / {\rm d}p_{\rm T}{\rm d}y$, 
${\rm d^2}\sigma^{\mu^\pm\leftarrow {\rm W/Z/\gamma^*}} / {\rm d}p_{\rm T}{\rm d}y$ 
and 
${\rm d^2}\sigma^{\mu^\pm\leftarrow {\rm J/\psi}} / {\rm d}p_{\rm T}{\rm d}y$
are 
the estimated $\pT$- and $y$-differential production cross sections of muons 
from primary charged-pion decays, primary charged-kaon decays, 
secondary (charged) pion and kaon decays, W and Z/$\gamma^*$ decays and 
J/$\psi$ decays, respectively.

The inclusive muon production cross section is determined according to:
\begin{linenomath}
\begin{equation}
{{{\rm d^2}\sigma^{\mu^\pm}} \over 
{{\rm d}p_{\rm T} {\rm d}y }} = 
{1 \over {A \times \epsilon}} \cdot
{{{\rm d^2}N^{\mu^\pm}} \over {{\rm d}p_{\rm T} {\rm d}y }} \cdot 
{1\over {{\mathcal L}_{\rm int}}},
\label{eq:CrossSecmu}
\end{equation}
\end{linenomath}
where  $A \times \epsilon$ is the product of acceptance and efficiency and ${\rm d^2}N^{\mu^\pm} / {\rm d}p_{\rm T}{\rm d}y $ is 
the measured $\pT$- and $y$-differential muon yield. 
The integrated luminosity ${\mathcal L}_{\rm int}$ is computed as 
$N_{\rm MSL(MSH)}$/$\sigma_{\rm MSL (MSH)}$, 
where $N_{\rm MSL(MSH)}$ and $\sigma_{\rm MSL (MSH)}$ are the number of 
MSL (MSH)-triggered events and the corresponding 
MSL (MSH)-trigger cross section. 
The latter is expressed as 
$\sigma_{\rm MSL (MSH)} = \sigma_{\rm T0}/F_{\rm MSL(MSH)}$, where 
$\sigma_{\rm T0}$ and $F_{\rm MSL(MSH)}$ are the visible cross section for T0 
measured with the van der Meer scan~\cite{ALICE-PUBLIC-2016-005} and the 
corresponding normalisation factor. The T0 cross section amounts to 
$\sigma_{\rm T0} = 21.6 \pm 0.4$~mb. The total systematic 
uncertainty of 2.1\% includes contributions from the T0 trigger cross section 
measurement and the stability of T0 during the data taking. The 
normalisation factors $F_{\rm MSL} = 34.30 \pm 0.05$ and 
$F_{\rm MSH} = 1370.9 \pm 2.2$ are the 
run-averaged ratio of T0 trigger rates corrected for pile-up to those 
of muon triggers (MSL or MSH) corrected by the 
fraction of events satisfying the event selection criteria. 
The quoted uncertainty is statistical, the systematic uncertainty being 
negligible (see Section~\ref{sec:syst}).

\begin{figure}[!t]
\begin{center}
\includegraphics[width=.7\columnwidth]{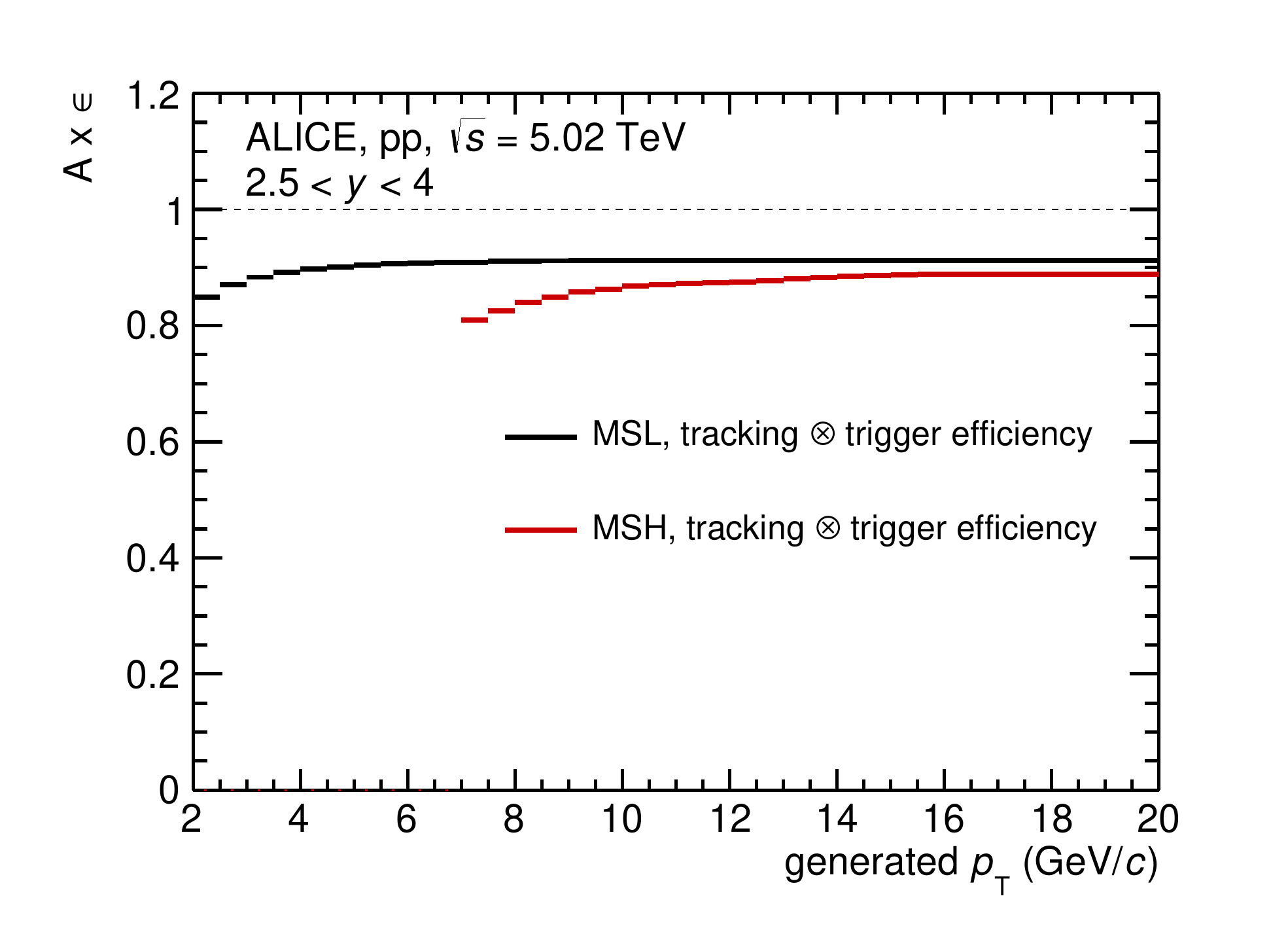}
\caption{Product of acceptance and efficiency as a function of generated $\pT$ 
estimated from a Monte Carlo simulation of muons from charm and beauty 
decays.}
\label{Fig:Eff}
\end{center}
\end{figure}

The measured $\pT$- and $y$-differential muon yields are corrected for the 
detector acceptance, tracking and trigger efficiencies ($A \times \epsilon$) 
using the same procedure as for previous 
analyses~\cite{Abelev:2012pi,Abelev:2012qh,Acharya:2017hdv}. 
The $A \times \epsilon$ corrections are evaluated from Monte Carlo 
simulations 
where muons from charm and beauty decays\footnote{It was verified that 
the $A \times \epsilon$ correction is the 
same for all muons, disregarding their origin, within systematic 
uncertainties, in the considered kinematic region.} 
are generated using 
the input $\pT$ and $y$ distributions predicted by FONLL 
calculations~\cite{Cacciari:2012ny}. 
These simulations are 
based on the GEANT3 transport code~\cite{Brun:1994aa} for the detector 
description and response, and include the time 
evolution of the detector configuration as well as alignment effects. 
The resulting $A \times \epsilon$ in MSL-triggered events is almost 
independent of $\pT$ and is about $90\%$ for $\pT > 4$~GeV/$c$, 
while in MSH-triggered events the 
$A \times \epsilon$ plateau is reached at higher $\pT$, 
about 15~GeV/$c$ (Fig.~\ref{Fig:Eff}).

The determination of the contribution of muons from charged pion and kaon 
decays, which 
dominates the background at low and intermediate 
$\pT$, is based on a data-tuned Monte Carlo cocktail. The procedure uses as 
inputs the $\pT$-differential mid-rapidity yields of charged pions and kaons 
per inelastic pp collision at $\sqrt{s}=5.02$~TeV,
$\lbrack{\rm d}^2 N^{\pi^\pm (K^\pm)} / {\rm d}p_{\rm T} {\rm d} y\rbrack_{{\rm mid-}y}$, 
resulting from an interpolation of 
data measured in 
pp collisions at $\sqrt s$ = 2.76 and 7 TeV, as described 
in~\cite{Abelev:2014laa,Adam:2015qaa,Adam:2016dau}. These reference $\pT$ spectra, measured up 
to $\pT$~= 20 GeV/$c$, are extrapolated to higher $\pT$ using a 
power-law fit to extend the $\pT$ coverage to the $\pT$ interval relevant
for the estimation of the contribution of decay muons up to $\pT=20$~GeV/$c$. 
Furthermore, the rapidity extrapolation of 
these distributions in a wider rapidity interval 
covering forward rapidities is performed according to:
\begin{linenomath}
\begin{equation}
\label{fextrap}
 \frac {{\rm d}^2 N^{\pi^\pm (K^\pm)}} 
{{\rm d}p_{\rm T} {\rm d} y} = 
F_{\rm extrap} (p_{\rm T}, y) \cdot
 \bigg\lbrack \frac {
{\rm d}^2 N^{\pi^\pm (K^\pm)}} {{\rm d}p_{\rm T} {\rm d} y}
\bigg\rbrack _{{\rm mid}-y},
\end{equation} 
\end{linenomath}
where $F_{\rm extrap} (p_{\rm T}, y)$ is the $\pT$-dependent rapidity 
extrapolation factor. 
The rapidity extrapolation is obtained from Monte Carlo simulations based on 
PYTHIA 6.4.25~\cite{Sjostrand:2006za} (Perugia-2011~\cite{Skands:2010ak}) 
and PHOJET~\cite{PhysRevD.52.1459} event 
generators. Furthermore, PYTHIA 8~\cite{Sjostrand:2014zea} simulations with 
various colour reconnection (CR) options 
("default MPI (Multi-Parton Interactions)", "new QCD" and "no CR") are 
employed 
to account for the $\pT$ dependence of the rapidity extrapolation and to 
estimate the related systematic uncertainty. It was also checked that 
PYTHIA 8~\cite{Sjostrand:2014zea} 
(Monash-2013~\cite{Skands:2014pea}) predictions give comparable results 
as PYTHIA 6 and PHOJET within uncertainties. 
Then, the $\pT$ and $y$ 
distributions of muons from the decay of charged pions and kaons are 
generated with a fast detector simulation of the decay kinematics and 
absorber effect, using as inputs the extrapolated 
primary charged pion and kaon spectra. 
The decay vertex of muons from charged pion and kaon decays is 
parameterised using either a single exponential for decays occurring before 
the front absorber ($z_{\rm v} \ge -90$~cm), or two exponentials for decays 
occurring inside the front absorber ($-503~{\rm cm}< z_{\rm v} < -90$~cm), 
in which 
case the first exponential represents the decay probability whereas the 
second corresponds to the hadron absorption probability. 
The fraction of reconstructed muons produced after the front absorber is 
negligible. Finally, the yields are converted into a cross section and subtracted from the inclusive 
muon distribution. 
The relative contributions of muons from primary charged pion decays and 
muons from primary charged kaon decays to inclusive muons are 
comparable. In the acceptance of the muon spectrometer, 
$2.5 < y <4$, 
the total contribution of muons from both charged pion and kaon decays 
decreases with increasing $\pT$ from about $39\%$ at $\pT=2$~GeV/$c$ 
down to $4\%$ at $\pT=20$~GeV/$c$. This background contamination depends 
also on 
$y$, in particular at low $\pT$ where it amounts to $47\%$ and $26\%$ in 
the 
rapidity intervals $2.5<y<2.8$ and $3.7<y< 4$, respectively.

The contribution of muons from secondary (charged) pion and kaon 
decays resulting from the interaction of light-charged hadrons with the 
material of the front absorber of the ALICE muon spectrometer 
is estimated by means of simulations 
using PYTHIA  6.425~\cite{Sjostrand:2006za} and the GEANT3 
transport code~\cite{Brun:1994aa}. This contribution affects the low $\pT$ 
region from $\pT$ = 2 GeV/$c$ up to about  $\pT$ = 5 GeV/$c$, only. 
The relative contribution with 
respect to inclusive muons decreases strongly with $\pT$, from about 4\% 
at $\pT$ = 2 GeV/$c$ to become smaller than 1\% at $\pT$ = 5 GeV/$c$. It also 
varies with rapidity, by decreasing down to about 3\% at $\pT$ = 2 GeV/$c$ 
in the interval $3.7 < y <4$. 

At high $\pT$, the W-boson decay muons and the dimuons from Z-boson decays 
and $\gamma^*$ decays (Drell-Yan process) are the main contributions to 
the background 
muon $\pT$ distribution. This background source is estimated with 
simulations using the POWHEG NLO 
event generator~\cite{Alioli:2008gx} paired with 
PYTHIA 6.425~\cite{Sjostrand:2006za} for parton shower simulation. 
These calculations use the CT10 PDFs~\cite{Lai:2010vv}. The relative contribution 
of muons from W and Z/$\gamma^*$ decays to the inclusive muon yield 
in $2.5 < y < 4$ is negligible for $\pT<12$~GeV/$c$ and increases 
significantly 
with $\pT$ from about $1\%$ at $\pT=12$~GeV/$c$ up to $12\%$ in 
$18<\pT< 20$~GeV/$c$. It also depends on rapidity and varies 
as a function of rapidity in the range $3\% - 6\%$ in the interval 
$14 < p_{\rm T} < 20$~GeV/$c$. 

The background component of muons from J/$\psi$ decays is estimated by 
means of a data-driven method similar 
to that implemented for the evaluation of muons from primary charged pion and 
kaon decays. The procedure uses the inclusive J/$\psi$ $\pT$- and 
$y$-differential 
cross sections measured by ALICE in the dimuon channel in the forward 
rapidity region ($2.5 < y <4$) at $\sqrt s$ = 5.02 TeV~\cite{Acharya:2017hjh}. 
The J/$\psi$ $\pT$ distribution being limited to the interval 
$p_{\rm T}<8$~GeV/$c$, it is fitted with the following function 
\begin{linenomath}
\begin{equation}
\label{jpsifunc}
f(p_{\rm T}) = C \cdot \frac {p_{\rm T}} {\bigg(1 + 
(\frac{p_{\rm T}}{p_0})^2\bigg)^n},
\end{equation} 
\end{linenomath}
where $C$, $p_0$ and $n$ are free parameters, and 
further extrapolated to higher $\pT$ values. The $y$ 
distribution is also extended in a wider range by means of a 
second-order polynomial function in order to avoid edge effects. 
Finally, the contribution of muons from J/$\psi$ decays is estimated 
with a simulation of the decay kinematics, using as inputs the 
extrapolated $\pT$ and $y$ production cross sections. 
As expected, this contamination is small compared to the other sources. 
The relative contribution with respect to the inclusive muon yield in the full 
acceptance of the muon spectrometer
is maximum at intermediate $\pT$ ($\pT$ $\sim 4-6$~GeV/$c$) where it amounts 
to about $4\%$ and decreases with increasing $\pT$ to become negligible for 
$p_{\rm T}>15$~GeV/$c$ (smaller than $1\%$). 
This background source exhibits a weak dependence on 
rapidity, with the maximum contribution at 
$\pT$ $\sim 4-6$~GeV/$c$ varying within $4\%-6\%$.

\begin{figure}[!t]
\begin{center}
\includegraphics[width=.7\columnwidth]{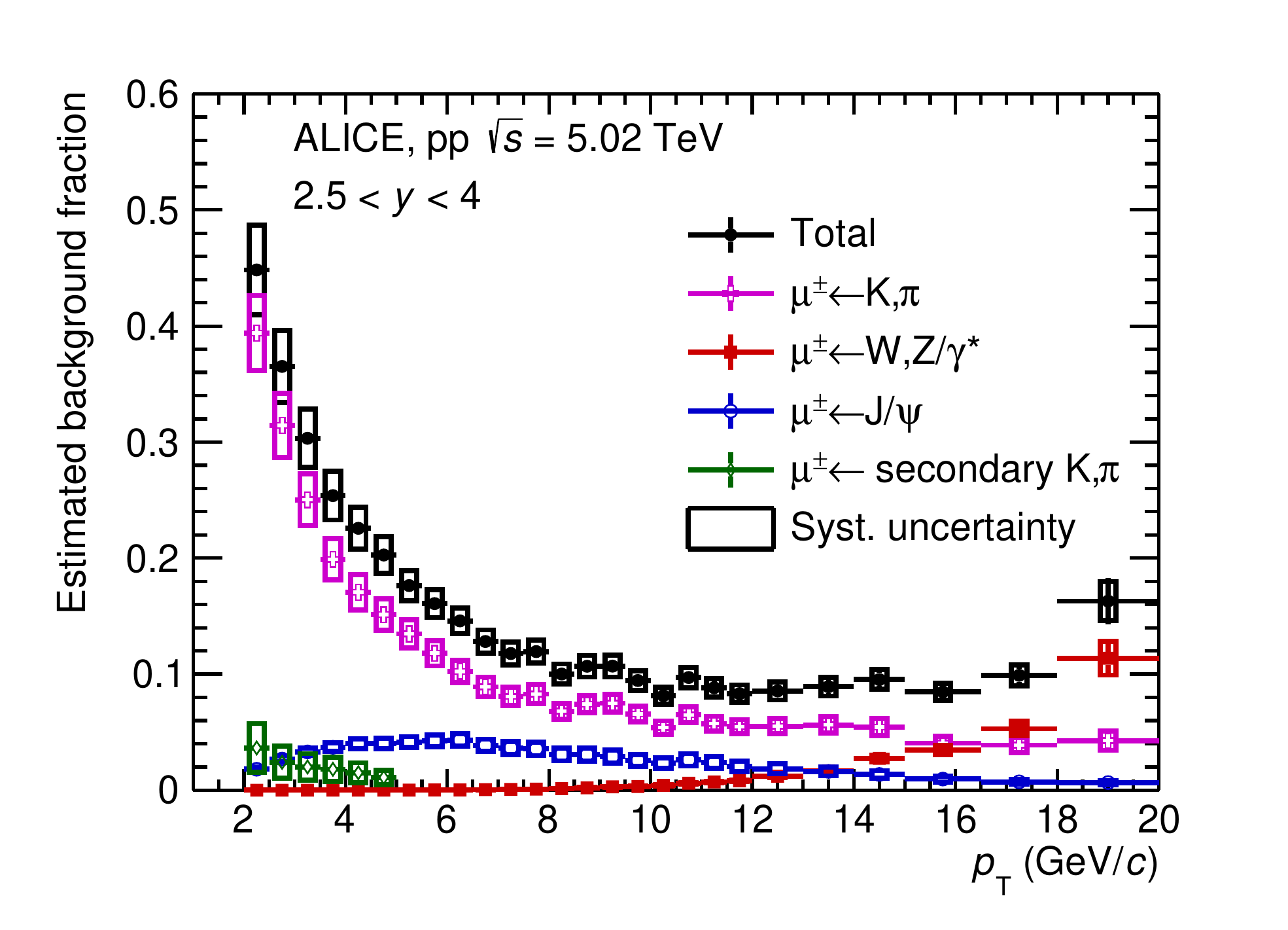}
\caption{Estimated background fractions with respect to inclusive muons as 
a function of $\pT$ for the rapidity interval $2.5 < y <4$ in pp collisions at 
$\sqrt s$ = 5.02 TeV. 
Statistical uncertainties (vertical bars) and systematic uncertainties (boxes) 
are shown. }
\label{Fig:bkg}
\end{center}
\end{figure}

Figure~\ref{Fig:bkg} summarises the estimated relative contribution of the 
various sources of background with respect to inclusive muons as a 
function of $\pT$ for the rapidity interval 
$2.5<y<4$, as well as the total background contamination. 
The vertical bars are the statistical uncertainties and the boxes 
are the systematic uncertainties on muon background sources 
that are discussed hereafter.

\subsection{Systematic uncertainties}\label{sec:syst}

Several sources of systematic uncertainty affecting the measurement of 
the $\pT$- and $y$-differential production cross section of muons from 
heavy-flavour hadron decays are evaluated. 
These are the systematic 
uncertainties on the inclusive muon yield, the estimated 
background sources and the determination of the integrated luminosity. 

The systematic uncertainty on the inclusive muon yield contains the following 
contributions. The systematic uncertainty on the muon tracking efficiency 
amounts to 0.5\% and is estimated by measuring the efficiency 
in data and Monte Carlo with a procedure that exploits the redundancy of the 
tracking chamber information~\cite{Abelev:2014ffa,Adam:2015isa}. 
The systematic uncertainty 
on the single muon trigger efficiency of 1.4\% (3.2\%) for 
MSL (MSH) trigger comes from the intrinsic efficiency of the trigger chambers 
and the response of the trigger 
algorithm. The first contribution is determined from the uncertainty on the 
trigger 
chamber efficiency measured in the data and applied to the simulations. 
The second one is estimated by comparing the 
$\pT$ dependence of the MSL and MSH trigger response function in data and 
Monte Carlo~\cite{Adam:2015isa}. A 0.5\% contribution related to the choice 
of the $\chi^2$ cut implemented for 
the matching between tracker and trigger tracks is also taken into account. 
The magnitude of these 
systematic uncertainties is approximately independent of the kinematics, 
in the region of interest. 
Finally, an additional contribution related to the tracking chamber resolution 
and alignment needs to be taken into account. 
The procedure employed for the estimation of this uncertainty is 
based on the one described in~\cite{Alice:2016wka}. It uses a Monte Carlo 
simulation modelling the tracker response of the 
muon spectrometer with a parameterisation of the tracking chamber resolution 
and systematic mis-alignment effects. 
The former is measured using the residual distance between the reconstructed 
tracks and their associated clusters. The 
latter is inferred by comparing the reconstructed $\pT$ distribution of 
positive and negative muons, which have opposite curvature in the dipole 
magnet field and thus opposite sensitivity to the mis-alignment. 
This parameterisation is 
tuned either on data or on the full Monte Carlo simulation. The comparison of 
the heavy-flavour decay muon $\pT$-differential distributions obtained with 
the two parameterisations gives an estimation of the systematic uncertainty. 
It is negligible 
for $p_{\rm T} < 7$~GeV/$c$ and then increases to about $15\%$ in 
the interval $18 < p_{\rm T} < 20$~GeV/$c$. 

The systematic uncertainty on the estimated yield of muons from primary 
charged $\pi$ 
($\rm K$) decays includes contributions from i) the measured mid-rapidity 
$\pT$ distributions of charged $\pi$ ($\rm K$) up to $\pT=20$~GeV/$c$ and 
their extrapolation to higher $\pT$, varying from about 
$7\%$ ($9\%$) to about $21\%$ ($22\%$) as a function of $\pT$, ii) the 
rapidity extrapolation of about $8.5\%$ ($6\%$) 
for muons from charged $\pi$ ($\rm K$), estimated by comparing the 
results with PYTHIA 6 and PHOJET generators iii) the $\pT$ dependence of the 
rapidity extrapolation, negligible for $p_{\rm T} < 4$~GeV/$c$ and 
increasing up to about $6\%$ ($3\%$) 
for charged $\rm \pi$ ($K$) decay muons, obtained by comparing the 
results with several colour reconnection options in PYTHIA 8 and iv) 
the simulation of hadronic interactions in the front absorber of 
about $4\%$ for both charged $\pi$ and $\rm K$ decay muons. The latter was 
estimated by comparing the $\pT$ 
distributions of muons from charged pion and kaon decays obtained in 
a fast detector 
simulation based on a parameterisation of the effects of the front 
absorber (Section~\ref{subsec:proc}) and a full simulation.
Combining these sources, a total systematic uncertainty ranging 
from about $11\%$ to $24\%$ as a function of $\pT$ is obtained, with 
approximately no dependence on the decay particle type. 
On the other hand, in order to account for the systematics associated to 
the transport code~\cite{Abelev:2012pi}, a conservative systematic 
uncertainty on the estimated 
yield of muons from secondary charged $\pi$ ($\rm K$) decays of 100\% is 
considered and the obtained difference between the upper and lower limits is 
further divided by $\sqrt{12}$, corresponding to one RMS of a uniform 
distribution.

The systematic uncertainty of the estimated yield of muons from 
W and Z/$\gamma^*$ decays is determined by considering the CT10 
PDF uncertainties. It amounts to about 8\% (7\%) for muons from W (Z/$\gamma^*$) decays
\footnote{A similar systematic uncertainty is also obtained by 
performing POWHEG 
simulations with CTEQ6M (NLO) PDF~\cite{Pumplin:2002vw}.}.   

The systematic uncertainty on the extracted yield of muons from 
J/$\psi$ originates from the measured J/$\psi$ $\pT$ and $y$ 
distributions and their extrapolation in a wider kinematic region, with a 
negligible effect on the extracted muon yield when using different functions 
for the rapidity extrapolation. 
This systematic uncertainty increases with increasing $\pT$ from about 
$10\%$ to $34\%$. 

The systematic uncertainty on the integrated luminosity reflects the 
2.1\% systematic uncertainty on the measurement of the T0 trigger cross 
section~\cite{ALICE-PUBLIC-2016-005}, the systematic uncertainty on 
the normalisation factor of 
muon-triggered events to the equivalent number of T0-triggered events based 
on the relative trigger rates being negligible. 
Indeed, compatible results are found when calculating the integrated 
luminosity for MSL (MSH) trigger by applying the corresponding trigger 
condition in the analysis of MB events, rather than using the 
relative trigger rates.

\begin{table}[!hbt] 
\centering
\begin{tabular}{c|c}
\hline
Source & Uncertainty vs $p_{\rm T}$\\ 
\hline 
Tracking efficiency & 0.5\% \\
Trigger efficiency & 1.4\% (3.2\%) for MSL (MSH) \\
Matching efficiency & 0.5\%  \\
Resolution and alignment & 0--15\% (negligible for $\pT$~$<$ 7 GeV/$c$)\\
Background subtraction $\mu \leftarrow \pi$ & $1-4.4$\%  \\
Background subtraction $\mu \leftarrow {\rm K}$ & $1-4.4$\%  \\
Background subtraction $\mu \leftarrow {\rm sec.\ \pi, K}$ & $0-4.3$\%  \\
Background subtraction $\mu \leftarrow W/Z/\gamma^*$ & $0-1.1$\%  \\
Background subtraction $\mu \leftarrow \rm J/\psi$ & $0-0.7$\%  \\
Integrated luminosity & 2.1\% \\
\hline
\end{tabular}
\caption{Summary of relative systematic uncertainties after propagation to 
the measurement of the $\pT$-differential cross section of muons from 
heavy-flavour hadron decays at forward rapidity 
($2.5 < y <4$). See the text for details. 
For the $\pT$-dependent uncertainties, the minimum and maximum values 
are given. 
They are shown for the lowest and highest $\pT$~interval 
with the exception of the light-hadron 
decay muon background, where this is the opposite trend, and of 
the background of muons from J/$\psi$ decays with the maximum value being 
reached for $4 < p_{\rm T} < 6$~GeV/$c$. The systematic uncertainty on the 
integrated luminosity is correlated as a function of $\pT$.}
\label{tab:SystUnc}
\end{table}

Table~\ref{tab:SystUnc} gives an overview of the systematic uncertainties 
assigned to the various contributions which enter in the measurement of the 
$\pT$-differential cross section of 
muons from heavy-flavour hadron decays in $2.5 < y<4$. The total systematic 
uncertainty is the quadratic sum of the sources listed in 
Tab.~\ref{tab:SystUnc}, with the exception of the 2.1\% contribution on 
the integrated luminosity which is fully correlated with $\pT$. It varies 
from about $2\%$ to $15\%$, the smaller (higher) value corresponding to 
$\pT=6.5~$GeV/$c$ ($18 < p_{\rm T} < 20$~GeV/$c$). In 
the high-$\pT$ region ($18 < p_{\rm T} < 20$~GeV/$c$), 
the main contribution comes 
from the uncertainty on tracking chamber resolution and alignment.

%% file: body/c04Results.tex

\begin{figure}[!t]
\begin{center}
\includegraphics[width=.65\columnwidth]{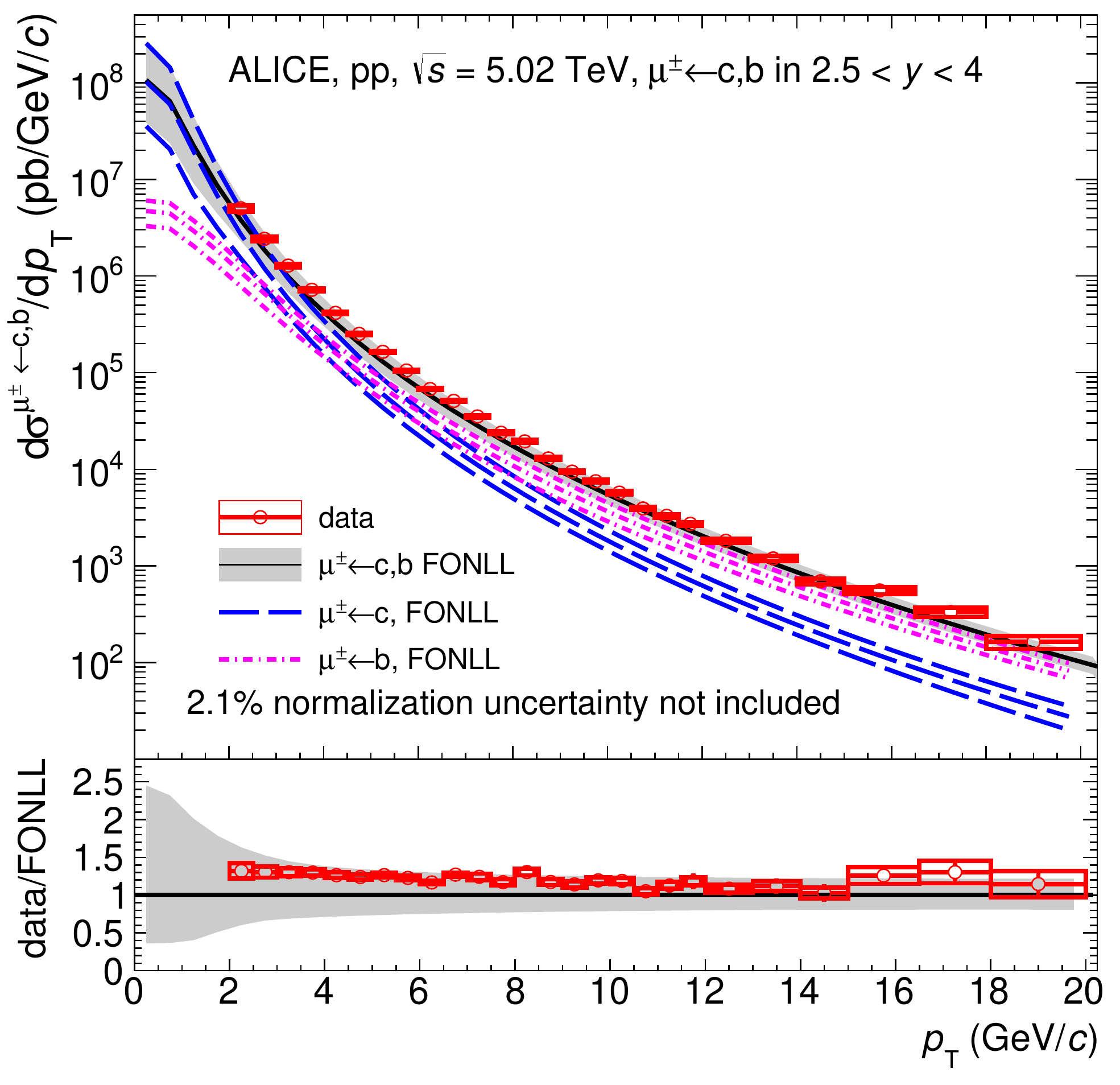}
\caption{$\pT$-differential production cross section of muons from 
heavy-flavour hadron decays at forward rapidity in pp collisions at 
$\sqrt s$ = 5.02 TeV. Statistical uncertainties (bars) and 
systematic uncertainties (boxes) are shown. The production cross section is 
compared with FONLL predictions~\cite{Cacciari:2012ny} (top). The ratio of 
the data to FONLL calculations is shown in the lower panel. See the text 
for details.}
\label{Fig:pTxsect}
\end{center}
\end{figure}

The $\pT$-differential cross section of muons from heavy-flavour hadron decays 
in $2.5 < y <4$ is presented in Fig.~\ref{Fig:pTxsect}. The vertical 
bars represent the statistical 
uncertainties and are smaller than the symbols in most $\pT$ bins, 
while the empty boxes correspond to the systematic uncertainties. 
The symbols are positioned 
horizontally at the centre of each $\pT$ bin and the horizontal bars represent 
the width of the $\pT$ interval. These conventions are applied from here 
onwards to the 
figures discussed in the following. 
The measurement is carried out in 
a wider $\pT$ range with respect to 
previous measurements in pp collisions~\cite{Abelev:2012pi,Abelev:2012qh}, 
the $\pT$ reach 
being extended from $\pT$ = 10 GeV/$c$ at $\sqrt s$ = 2.76 TeV ($\pT$ = 
12 GeV/$c$ at 
$\sqrt s$ = 7 TeV) to $\pT$ = 20 GeV/$c$ by 
using MSL and MSH triggers. The total uncertainties (quadratic sum of 
statistical and systematic uncertainties) are reduced by a factor of about $2-4$ 
with respect 
to previous measurements. These improvements have various sources: i) better 
understanding of the detector response, ii) new data-driven strategy for the 
estimation of the contribution of muons from light-hadron decays, iii) larger 
integrated luminosity and iv) use of a high-$\pT$ trigger. The measured production cross section 
(Fig.~\ref{Fig:pTxsect}, upper panel) is compared with FONLL 
predictions. The FONLL calculations~\cite{Cacciari:2012ny, Cacciari:2015fta} include the non-per\-tur\-ba\-ti\-ve fragmentation into open heavy-flavour hadrons and their decay into final-state leptons. As described in~\cite{Cacciari:2005rk}, the production of leptons from charm- and beauty-hadron decays is controlled by measured decay spectra and branching ratios. These predictions which use 
the CTEQ6.6 PDFs~\cite{Nadolsky:2008zw} are represented with a black curve 
and a shaded band 
for the systematic uncertainty. 
The latter contains the uncertainties on the renormalization and factorization 
scales, on quark masses as well as on the PDFs.
The FONLL predictions are also displayed for muons 
coming from charm and beauty decays, separately. The latter contribution 
includes 
direct decays and decays via D-hadron decays. 
The FONLL predictions are compatible with data within the experimental 
and theoretical uncertainties. However, one can notice that the central 
values of FONLL predictions systematically underestimate the measured 
production cross section at low and intermediate $\pT$, i.e. up to 
$\pT$ $\simeq$ 8 GeV/$c$. 
This is also illustrated in the bottom panel of Fig.~\ref{Fig:pTxsect}, which 
shows the ratio between the measured production cross section and the FONLL 
calculations. This ratio is about 1.3 for $2 < p_{\rm T} < 8$~GeV/$c$ and then 
decreases with increasing $\pT$ to tend towards unity in the 
high $\pT$ region ($p_{\rm T} > 11-12$~GeV/$c$). 
Qualitatively, this behaviour was also reported at forward rapidity for muons 
from 
heavy-flavour hadron decays in previous 
analyses~\cite{Abelev:2012pi,Abelev:2012qh} and for D mesons measured 
in pp collisions at $\sqrt s$ = 5.02 and 13 TeV with the LHCb 
detector~\cite{Aaij:2016jht,Aaij:2015bpa}, as well as at mid-rapidity 
for D mesons and electrons from B-hadron and 
heavy-flavour hadron decays measured in pp collisions at $\sqrt s$ = 2.76 
and 7 TeV with ALICE~\cite{Abelev:2014hla,Abelev:2012sca,
Abelev:2014gla,Abelev:2012vra,Acharya:2017jgo}.

The measurement described here provides the baseline for 
the study of QCD matter created in Pb--Pb collisions at the same 
centre-of-mass energy and in Xe--Xe collisions at $\sqrt{s_{\rm NN}}$ 
= 5.44 TeV by applying a pQCD-driven energy scaling based on FONLL 
calculations~\cite{Averbeck:2011ga}. 
\begin{figure}[!hbt]
\begin{center}
\includegraphics[width=.49\columnwidth]{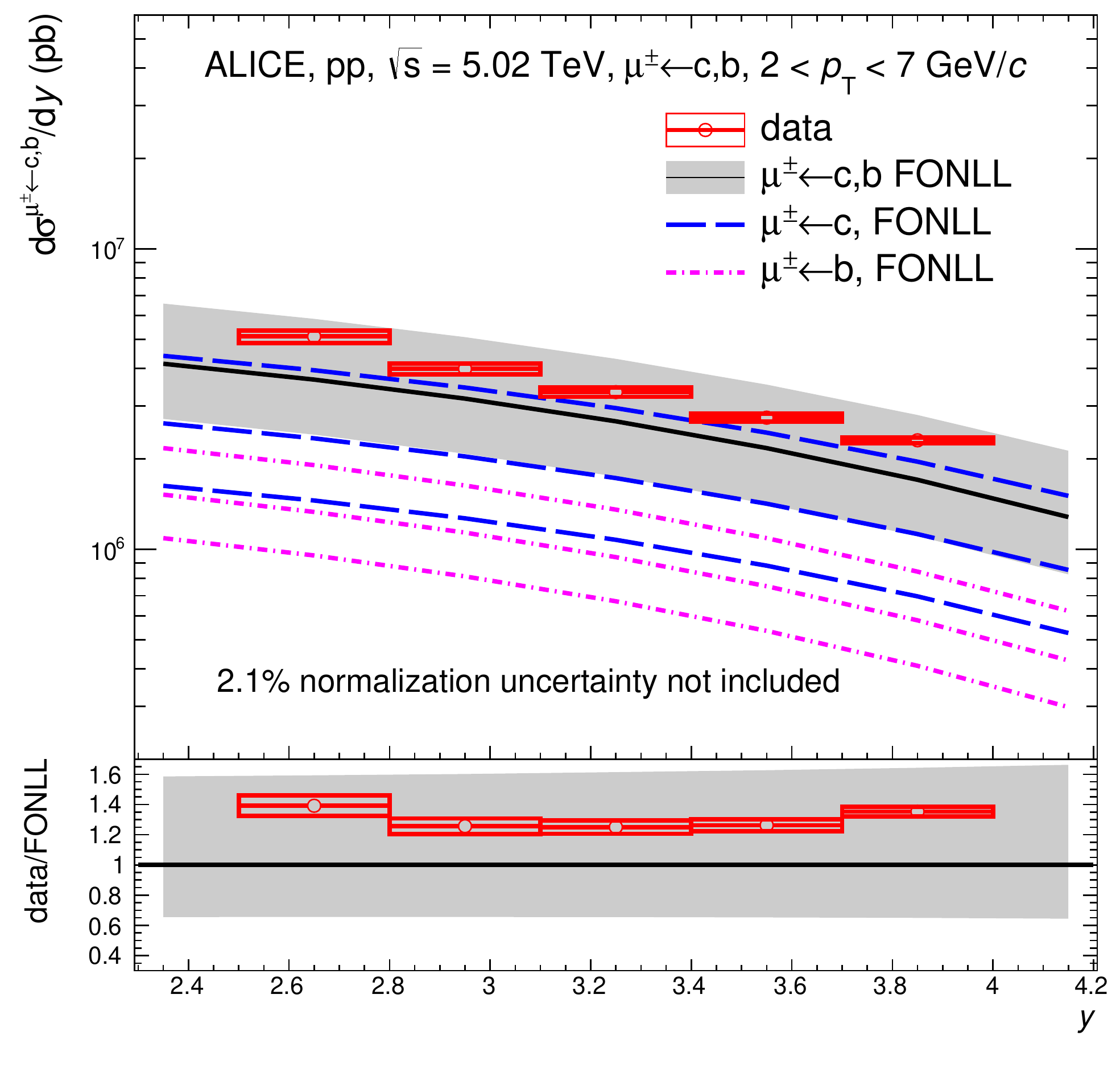}
\includegraphics[width=.49\columnwidth]{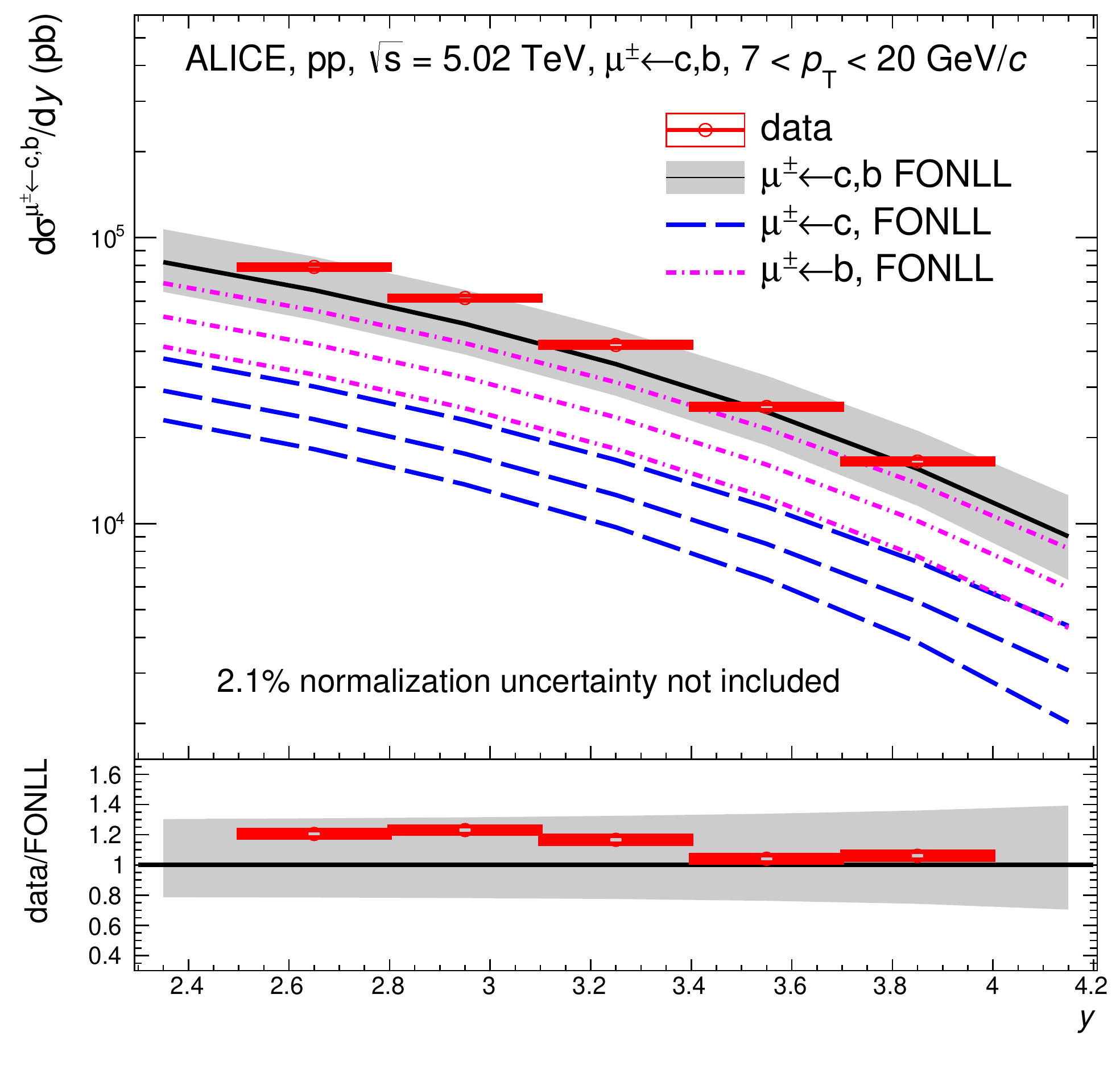}
\caption{Production cross section of muons from 
heavy-flavour hadron decays as a function of rapidity in pp collisions at 
$\sqrt s$ = 5.02 TeV for the $\pT$~intervals 
$2 < p_{\rm T} < 7$~GeV/$c$ (left) 
and $7 < p_{\rm T} < 20$~GeV/$c$ (right). Statistical uncertainties (bars, 
smaller than symbols) and 
systematic uncertainties (boxes) are drawn. The production cross sections are
compared with FONLL predictions~\cite{Cacciari:2012ny} (top). The ratios of 
the data to FONLL calculations are shown in the lower panels. See the text 
for details.}
\label{Fig:yxsect}
\end{center}
\end{figure}

\begin{figure}[!hbt]
\begin{center}
\includegraphics[width=.63\columnwidth]{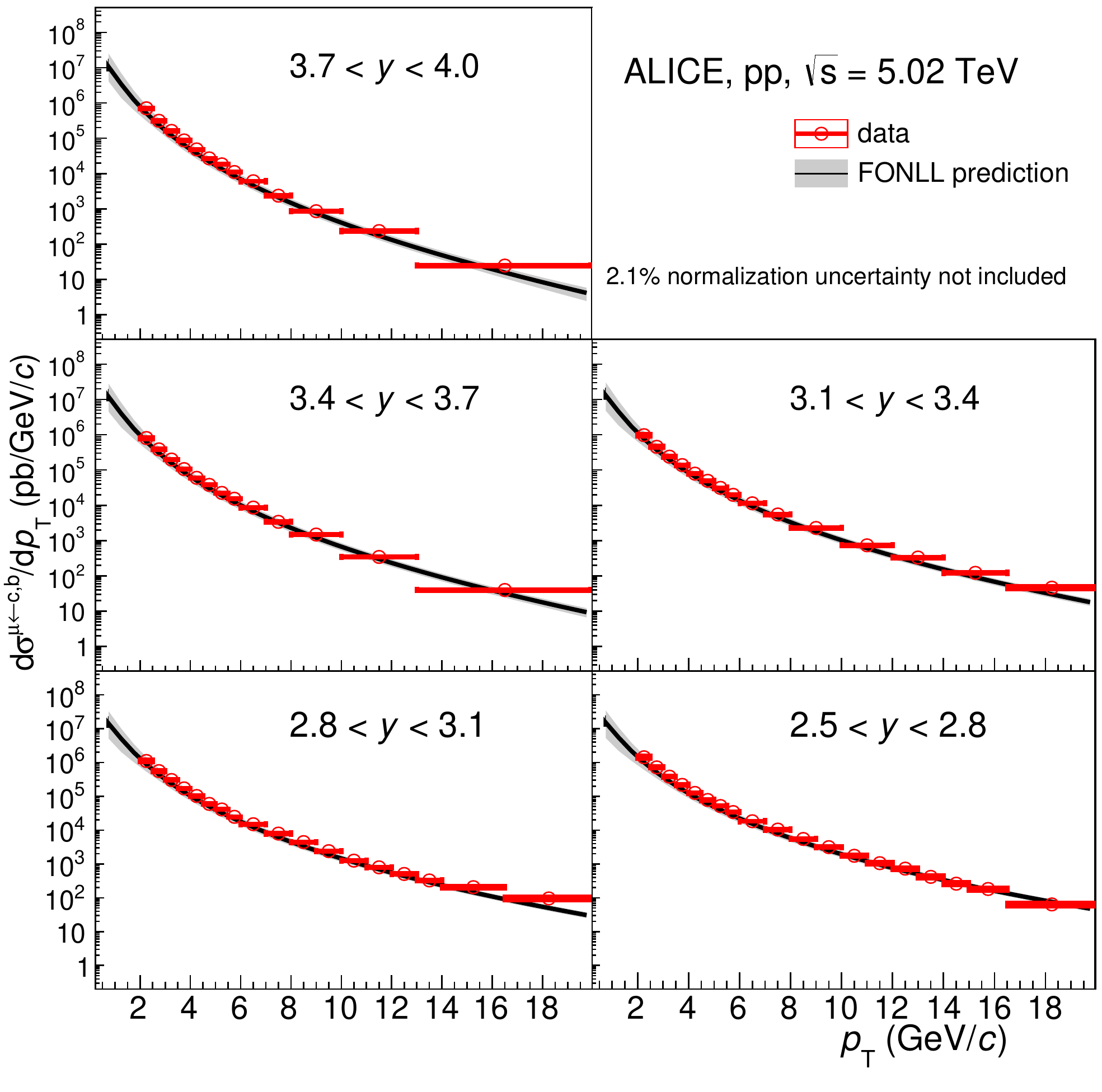}
\includegraphics[width=.63\columnwidth]{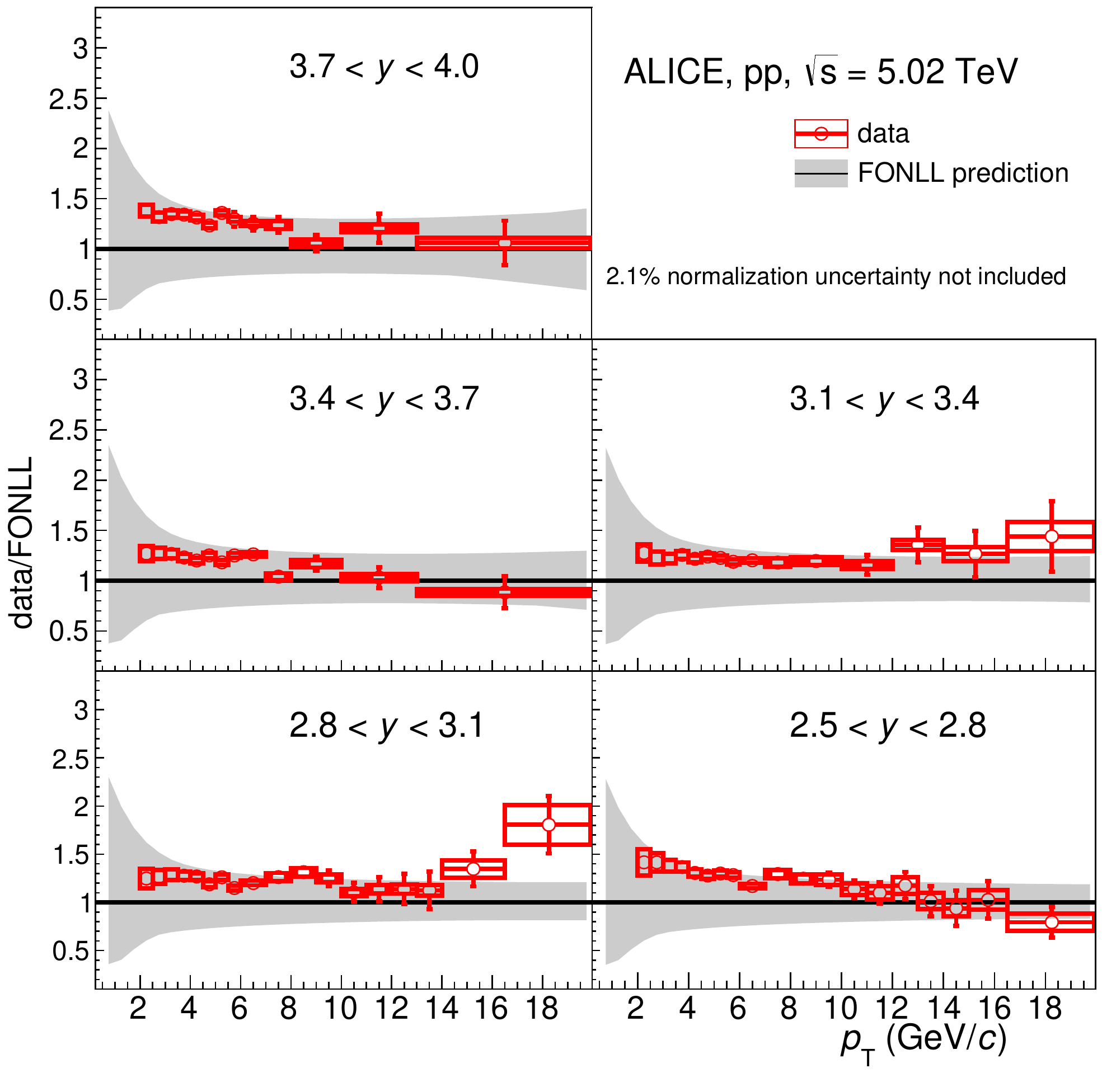}
\caption{Upper panel: $\pT$-differential production cross section of muons 
from 
heavy-flavour hadron decays for five rapidity intervals in the 
range $2.5 < y < 4$ in pp collisions at 
$\sqrt s$ = 5.02 TeV. Statistical uncertainties (bars) and 
systematic uncertainties (boxes) are shown. The production cross sections are 
compared with FONLL predictions~\cite{Cacciari:2012ny}. Bottom panel: ratios 
of the data to FONLL calculations. See the text 
for details.}
\label{Fig:ptyxsect}
\end{center}
\end{figure}
The $\pT$-integrated production cross section of muons from heavy-flavour 
hadron decays is also studied as a function of rapidity for the $\pT$ 
intervals 
$2 < p_{\rm T} < 7$~GeV/$c$ and $7 < p_{\rm T} < 20$~GeV/$c$, as shown 
in left and right panels of Fig.~\ref{Fig:yxsect}, respectively. The ratios 
between 
data and FONLL predictions are depicted in the bottom panels. 
The two measurements are 
consistent with FONLL predictions. As in the case of the 
$\pT$-differential 
production cross section, the data lie in the upper part of the FONLL 
predictions. In the interval 
$2 < p_{\rm T} < 7$~GeV/$c$, muons from heavy-flavour hadron decays originate 
predominantly from charmed hadrons, while in the higher $\pT$ region, muons 
from beauty-hadron decays take over from charm as the dominant source. 
One notices that in the 
higher $\pT$ interval, the agreement between data and 
the central values of FONLL calculations is better. 
The ratio of the measured production 
cross section to FONLL predictions is in the range $\sim 1-1.2$, 
depending on the rapidity region.

The statistics collected with muon triggers allows us to perform measurements 
of the $\pT$-differential cross section in five $y$ intervals in the range 
$2.5 < y <4$. The results and comparisons with FONLL are presented in 
Fig.~\ref{Fig:ptyxsect}, upper panel. The corresponding ratios between data 
and FONLL calculations are also displayed in Fig.~\ref{Fig:ptyxsect}, 
lower panel. The data and FONLL 
exhibit a good agreement within experimental and theoretical 
uncertainties, the former being 
systematically higher than the model calculations with some fluctuations at 
high $\pT$.

\begin{figure}[!hbt]
\begin{center}
\includegraphics[width=.7\columnwidth]{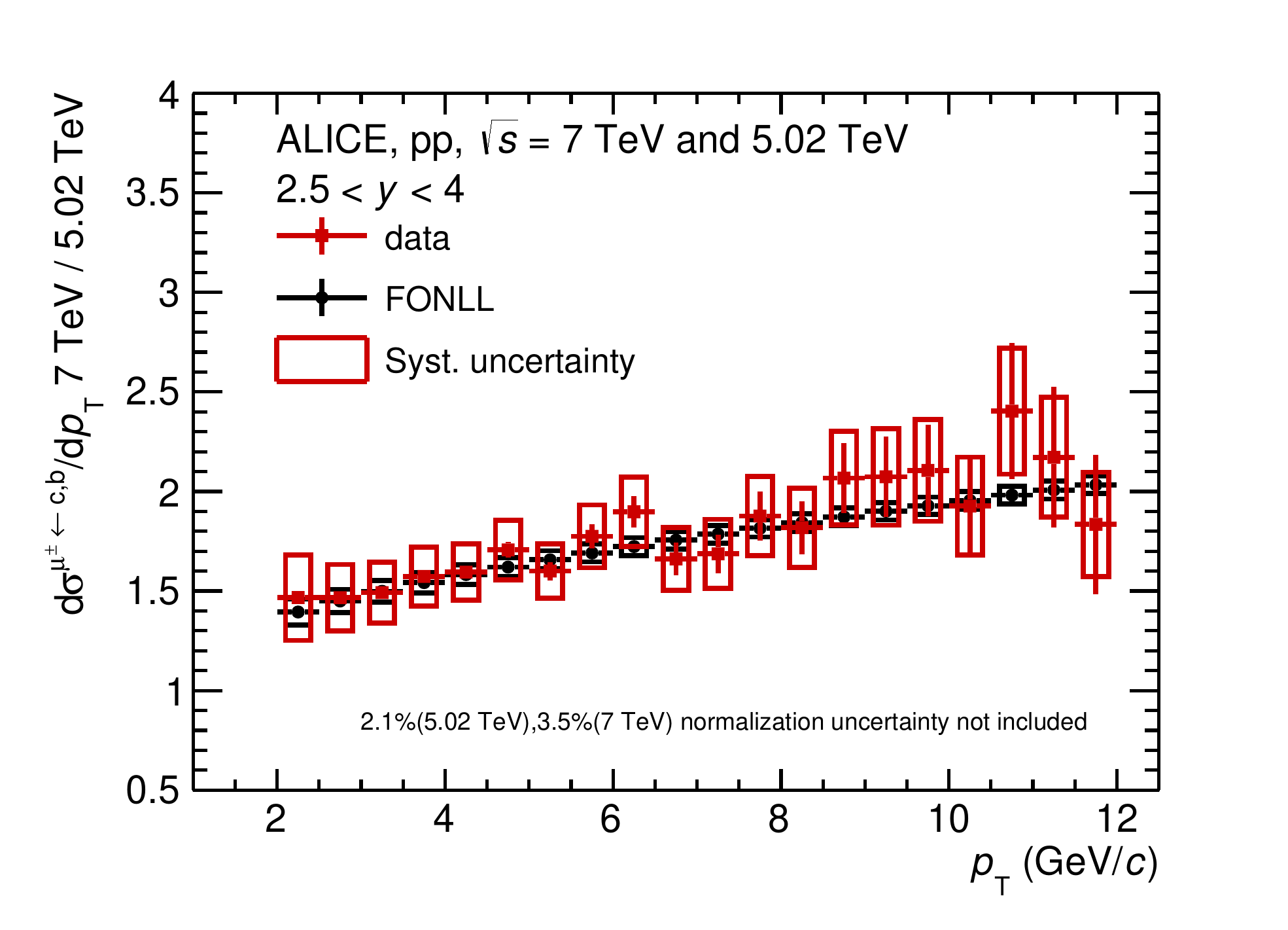}
\caption{Ratio of the $\pT$-differential production cross section of muons 
from heavy-flavour hadron decays at forward rapidity in pp collisions at 
$\sqrt s$ = 7 TeV to that at $\sqrt s$ = 5.02 TeV. 
Statistical uncertainties (bars) and 
systematic uncertainties (boxes) are shown. The normalisation uncertainty 
contains the uncertainties on the luminosity 
at the two centre-of mass energies. The ratio is 
compared with FONLL predictions~\cite{Cacciari:2012ny}. See the text 
for details.}
\label{Fig:ratio1}
\end{center}
\end{figure}

The ratio of open heavy-flavour production cross sections between 
different centre-of-mass energies is considered as a powerful observable for 
sensitive tests of pQCD-based calculations and to constrain gluon PDF at 
forward rapidity~\cite{Cacciari:2015fta}. While the absolute production cross 
sections as predicted by FONLL are associated with large systematic 
uncertainties, dominated by the scale uncertainties, the ratios of 
production cross sections at different centre-of-mass energies are predicted 
with a better accuracy~\cite{Cacciari:2015fta}. The ratio of the measured 
$\pT$-differential cross section of muons from heavy-flavour hadron 
decays in pp collisions at $\sqrt s$ = 7 TeV to that at $\sqrt s$ = 5.02 TeV 
in the rapidity interval $2.5 < y <4$ is reported in Fig.~\ref{Fig:ratio1}.  
The systematic uncertainties between the two measurements are considered as 
uncorrelated when forming the ratio and the main contribution comes from the 
measurement at $\sqrt s$ = 7 TeV. 
The ratio exhibits a smooth increase with increasing $\pT$ from 
about 1.5 ($\pT$~= 2 GeV/$c$) to 1.8 ($\pT$~= 12 GeV/$c$). 
The data are compared with FONLL 
predictions~\cite{Cacciari:2012ny}. The measured ratio is well 
reproduced by FONLL calculations. 

\begin{figure}[!hbt]
\begin{center}
\includegraphics[width=.7\columnwidth]{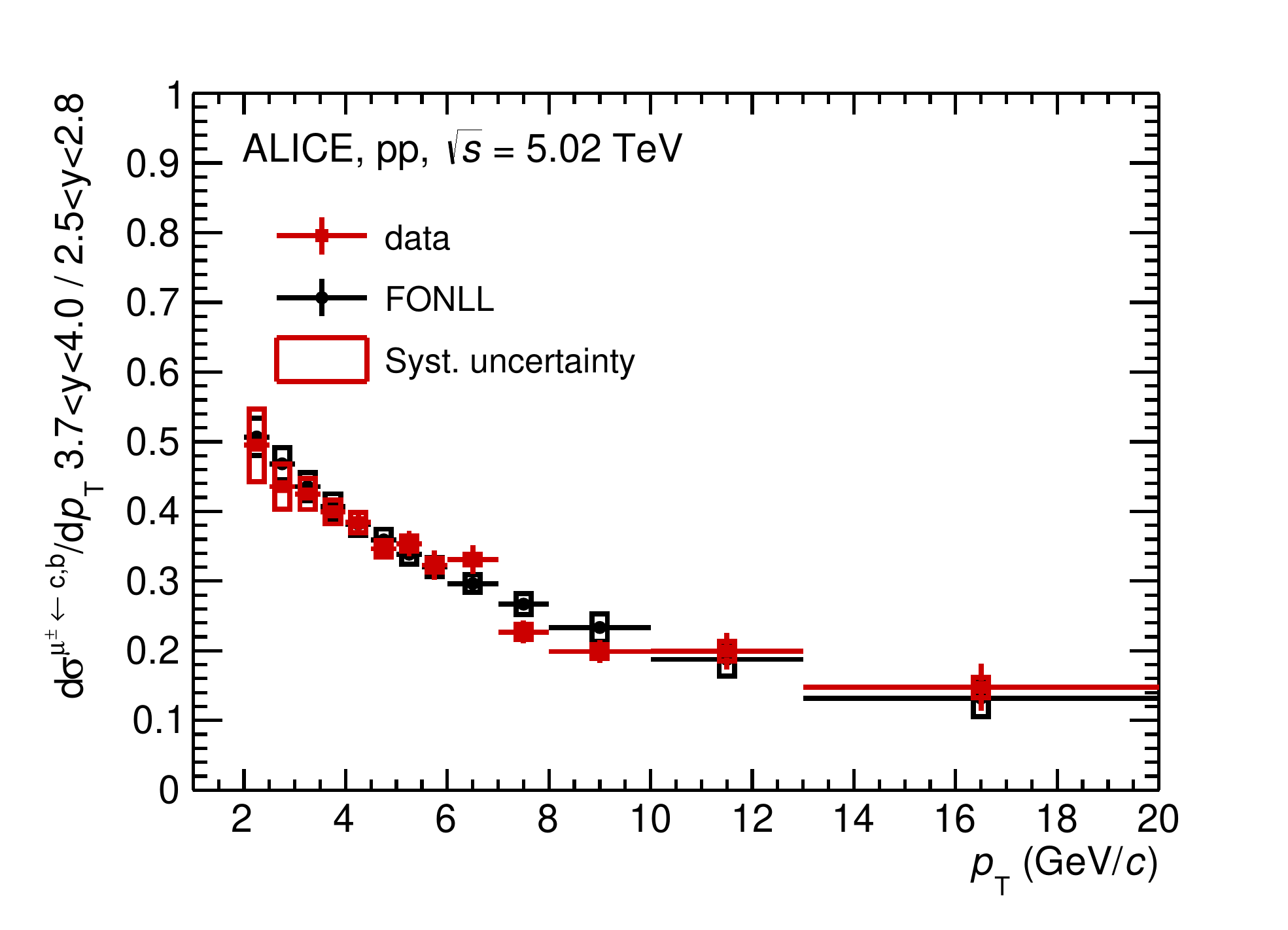}
\caption{Ratio of the $\pT$-differential production cross section of muons 
from heavy-flavour hadron decays in $3.7 < y < 4$ to that in $2.5 < y <2.8$ 
in pp collisions at $\sqrt s$ = 5.02 TeV. 
Statistical uncertainties (bars) and 
systematic uncertainties (boxes) are shown. The ratio is 
compared with FONLL predictions~\cite{Cacciari:2012ny}. See the text 
for details.}
\label{Fig:ratio2}
\end{center}
\end{figure}

A reduction of the systematic uncertainty on the FONLL predictions is also 
expected from the ratio of open heavy-flavour 
cross sections between different rapidity intervals, which could provide 
constraints on the gluon PDF at small Bjorken-$x$ values. This ratio, computed for heavy-flavour hadron decay muons between the two 
extreme rapidity intervals, i.e. $2.5 < y < 2.8$ and $3.7 < y < 4$, is 
presented in Fig.~\ref{Fig:ratio2}. When forming the ratio, the systematic 
uncertainty on integrated luminosity is correlated, while the 
systematic uncertainty on tracking chamber resolution and alignment is 
partially correlated. 
The other sources of systematic uncertainties are treated as uncorrelated. 
The ratio decreases significantly 
with increasing $\pT$ from about 0.5 down to 0.15. 
The measured ratio is compared with FONLL 
predictions, which describe the data within their uncertainties. 

%% file: body/c05Conclusion.tex

In summary, the production of muons from heavy-flavour hadron decays has been 
measured in the forward rapidity region as a function of $\pT$~and $y$ 
in pp collisions at $\sqrt {s}$ = 5.02 TeV with the ALICE 
detector at the CERN LHC. As compared to previously published measurements,
the present results have an extended $\pT$ coverage, 
$2 < p_{\rm T} < 20$~GeV/$c$, and a better precision with 
the total uncertainties reduced by 
a factor of about $2-4$, depending on $\pT$. The results provide the crucial 
reference for the study of the effects of the hot and dense matter on 
the production of muons from heavy-flavour hadron decays in Pb--Pb 
collisions at the same centre-of-mass energy. 
The measurements of the differential production cross sections are found 
to be in agreement with FONLL 
predictions over the full $\pT$ range, even though the central values of 
FONLL appear to underestimate the heavy-flavour hadron decay muon production. 
The $\pT$-differential ratios of the production cross section between 
$\sqrt s$ = 7 TeV and $\sqrt s$ = 5.02 TeV and between two rapidity intervals 
within $2.5 < y <4$ are well described by FONLL 
calculations.

%% file: fa_2019-03-23.tex

The ALICE Collaboration would like to thank all its engineers and technicians for their invaluable contributions to the construction of the experiment and the CERN accelerator teams for the outstanding performance of the LHC complex.
The ALICE Collaboration gratefully acknowledges the resources and support provided by all Grid centres and the Worldwide LHC Computing Grid (WLCG) collaboration.
The ALICE Collaboration acknowledges the following funding agencies for their support in building and running the ALICE detector:
A. I. Alikhanyan National Science Laboratory (Yerevan Physics Institute) Foundation (ANSL), State Committee of Science and World Federation of Scientists (WFS), Armenia;
Austrian Academy of Sciences, Austrian Science Fund (FWF): [M 2467-N36] and Nationalstiftung f\"{u}r Forschung, Technologie und Entwicklung, Austria;
Ministry of Communications and High Technologies, National Nuclear Research Center, Azerbaijan;
Conselho Nacional de Desenvolvimento Cient\'{\i}fico e Tecnol\'{o}gico (CNPq), Universidade Federal do Rio Grande do Sul (UFRGS), Financiadora de Estudos e Projetos (Finep) and Funda\c{c}\~{a}o de Amparo \`{a} Pesquisa do Estado de S\~{a}o Paulo (FAPESP), Brazil;
Ministry of Science \& Technology of China (MSTC), National Natural Science Foundation of China (NSFC) and Ministry of Education of China (MOEC) , China;
Croatian Science Foundation and Ministry of Science and Education, Croatia;
Centro de Aplicaciones Tecnol\'{o}gicas y Desarrollo Nuclear (CEADEN), Cubaenerg\'{\i}a, Cuba;
Ministry of Education, Youth and Sports of the Czech Republic, Czech Republic;
The Danish Council for Independent Research | Natural Sciences, the Carlsberg Foundation and Danish National Research Foundation (DNRF), Denmark;
Helsinki Institute of Physics (HIP), Finland;
Commissariat \`{a} l'Energie Atomique (CEA), Institut National de Physique Nucl\'{e}aire et de Physique des Particules (IN2P3) and Centre National de la Recherche Scientifique (CNRS) and R\'{e}gion des  Pays de la Loire, France;
Bundesministerium f\"{u}r Bildung und Forschung (BMBF) and GSI Helmholtzzentrum f\"{u}r Schwerionenforschung GmbH, Germany;
General Secretariat for Research and Technology, Ministry of Education, Research and Religions, Greece;
National Research, Development and Innovation Office, Hungary;
Department of Atomic Energy Government of India (DAE), Department of Science and Technology, Government of India (DST), University Grants Commission, Government of India (UGC) and Council of Scientific and Industrial Research (CSIR), India;
Indonesian Institute of Science, Indonesia;
Centro Fermi - Museo Storico della Fisica e Centro Studi e Ricerche Enrico Fermi and Istituto Nazionale di Fisica Nucleare (INFN), Italy;
Institute for Innovative Science and Technology , Nagasaki Institute of Applied Science (IIST), Japan Society for the Promotion of Science (JSPS) KAKENHI and Japanese Ministry of Education, Culture, Sports, Science and Technology (MEXT), Japan;
Consejo Nacional de Ciencia (CONACYT) y Tecnolog\'{i}a, through Fondo de Cooperaci\'{o}n Internacional en Ciencia y Tecnolog\'{i}a (FONCICYT) and Direcci\'{o}n General de Asuntos del Personal Academico (DGAPA), Mexico;
Nederlandse Organisatie voor Wetenschappelijk Onderzoek (NWO), Netherlands;
The Research Council of Norway, Norway;
Commission on Science and Technology for Sustainable Development in the South (COMSATS), Pakistan;
Pontificia Universidad Cat\'{o}lica del Per\'{u}, Peru;
Ministry of Science and Higher Education and National Science Centre, Poland;
Korea Institute of Science and Technology Information and National Research Foundation of Korea (NRF), Republic of Korea;
Ministry of Education and Scientific Research, Institute of Atomic Physics and Ministry of Research and Innovation and Institute of Atomic Physics, Romania;
Joint Institute for Nuclear Research (JINR), Ministry of Education and Science of the Russian Federation, National Research Centre Kurchatov Institute, Russian Science Foundation and Russian Foundation for Basic Research, Russia;
Ministry of Education, Science, Research and Sport of the Slovak Republic, Slovakia;
National Research Foundation of South Africa, South Africa;
Swedish Research Council (VR) and Knut \& Alice Wallenberg Foundation (KAW), Sweden;
European Organization for Nuclear Research, Switzerland;
National Science and Technology Development Agency (NSDTA), Suranaree University of Technology (SUT) and Office of the Higher Education Commission under NRU project of Thailand, Thailand;
Turkish Atomic Energy Agency (TAEK), Turkey;
National Academy of  Sciences of Ukraine, Ukraine;
Science and Technology Facilities Council (STFC), United Kingdom;
National Science Foundation of the United States of America (NSF) and United States Department of Energy, Office of Nuclear Physics (DOE NP), United States of America.

%% file: 2019-03-23-Alice_Authorlist_2019-Mar-23.tex

\begingroup
\small
\begin{flushleft}
S.~Acharya\Irefn{org141}\And 
D.~Adamov\'{a}\Irefn{org93}\And 
S.P.~Adhya\Irefn{org141}\And 
A.~Adler\Irefn{org74}\And 
J.~Adolfsson\Irefn{org80}\And 
M.M.~Aggarwal\Irefn{org98}\And 
G.~Aglieri Rinella\Irefn{org34}\And 
M.~Agnello\Irefn{org31}\And 
N.~Agrawal\Irefn{org10}\And 
Z.~Ahammed\Irefn{org141}
% ...
\And
\renewcommand\labelenumi{\textsuperscript{\theenumi}~}

\section*{Affiliation notes}
\renewcommand\theenumi{\roman{enumi}}
\begin{Authlist}
\item \Adef{org*}Deceased
\item \Adef{orgI}Dipartimento DET del Politecnico di Torino, Turin, Italy
\item \Adef{orgII}M.V. Lomonosov Moscow State University, D.V. Skobeltsyn Institute of Nuclear, Physics, Moscow, Russia
\item \Adef{orgIII}Department of Applied Physics, Aligarh Muslim University, Aligarh, India
\item \Adef{orgIV}Institute of Theoretical Physics, University of Wroclaw, Poland
\end{Authlist}

\section*{Collaboration Institutes}
\renewcommand\theenumi{\arabic{enumi}~}
\begin{Authlist}
\item \Idef{org1}A.I. Alikhanyan National Science Laboratory (Yerevan Physics Institute) Foundation, Yerevan, Armenia
\item \Idef{org2}Bogolyubov Institute for Theoretical Physics, National Academy of Sciences of Ukraine, Kiev, Ukraine
\item \Idef{org3}Bose Institute, Department of Physics  and Centre for Astroparticle Physics and Space Science (CAPSS), Kolkata, India
\item \Idef{org4}Budker Institute for Nuclear Physics, Novosibirsk, Russia
\item \Idef{org5}California Polytechnic State University, San Luis Obispo, California, United States
\item \Idef{org6}Central China Normal University, Wuhan, China
\item \Idef{org7}Centre de Calcul de l'IN2P3, Villeurbanne, Lyon, France
\item \Idef{org8}Centro de Aplicaciones Tecnol\'{o}gicas y Desarrollo Nuclear (CEADEN), Havana, Cuba
\item \Idef{org9}Centro de Investigaci\'{o}n y de Estudios Avanzados (CINVESTAV), Mexico City and M\'{e}rida, Mexico
\item \Idef{org10}Centro Fermi - Museo Storico della Fisica e Centro Studi e Ricerche ``Enrico Fermi', Rome, Italy
\item \Idef{org11}Chicago State University, Chicago, Illinois, United States
\item \Idef{org12}China Institute of Atomic Energy, Beijing, China
\item \Idef{org13}Chonbuk National University, Jeonju, Republic of Korea
\item \Idef{org14}Comenius University Bratislava, Faculty of Mathematics, Physics and Informatics, Bratislava, Slovakia
\item \Idef{org15}COMSATS University Islamabad, Islamabad, Pakistan
\item \Idef{org16}Creighton University, Omaha, Nebraska, United States
\item \Idef{org17}Department of Physics, Aligarh Muslim University, Aligarh, India
\item \Idef{org18}Department of Physics, Pusan National University, Pusan, Republic of Korea
\item \Idef{org19}Department of Physics, Sejong University, Seoul, Republic of Korea
\item \Idef{org20}Department of Physics, University of California, Berkeley, California, United States
\item \Idef{org21}Department of Physics, University of Oslo, Oslo, Norway
\item \Idef{org22}Department of Physics and Technology, University of Bergen, Bergen, Norway
\item \Idef{org23}Dipartimento di Fisica dell'Universit\`{a} 'La Sapienza' and Sezione INFN, Rome, Italy
\item \Idef{org24}Dipartimento di Fisica dell'Universit\`{a} and Sezione INFN, Cagliari, Italy
\item \Idef{org25}Dipartimento di Fisica dell'Universit\`{a} and Sezione INFN, Trieste, Italy
\item \Idef{org26}Dipartimento di Fisica dell'Universit\`{a} and Sezione INFN, Turin, Italy
\item \Idef{org27}Dipartimento di Fisica e Astronomia dell'Universit\`{a} and Sezione INFN, Bologna, Italy
\item \Idef{org28}Dipartimento di Fisica e Astronomia dell'Universit\`{a} and Sezione INFN, Catania, Italy
\item \Idef{org29}Dipartimento di Fisica e Astronomia dell'Universit\`{a} and Sezione INFN, Padova, Italy
\item \Idef{org30}Dipartimento di Fisica `E.R.~Caianiello' dell'Universit\`{a} and Gruppo Collegato INFN, Salerno, Italy
\item \Idef{org31}Dipartimento DISAT del Politecnico and Sezione INFN, Turin, Italy
\item \Idef{org32}Dipartimento di Scienze e Innovazione Tecnologica dell'Universit\`{a} del Piemonte Orientale and INFN Sezione di Torino, Alessandria, Italy
\item \Idef{org33}Dipartimento Interateneo di Fisica `M.~Merlin' and Sezione INFN, Bari, Italy
\item \Idef{org34}European Organization for Nuclear Research (CERN), Geneva, Switzerland
\item \Idef{org35}Faculty of Electrical Engineering, Mechanical Engineering and Naval Architecture, University of Split, Split, Croatia
\item \Idef{org36}Faculty of Engineering and Science, Western Norway University of Applied Sciences, Bergen, Norway
\item \Idef{org37}Faculty of Nuclear Sciences and Physical Engineering, Czech Technical University in Prague, Prague, Czech Republic
\item \Idef{org38}Faculty of Science, P.J.~\v{S}af\'{a}rik University, Ko\v{s}ice, Slovakia
\item \Idef{org39}Frankfurt Institute for Advanced Studies, Johann Wolfgang Goethe-Universit\"{a}t Frankfurt, Frankfurt, Germany
\item \Idef{org40}Gangneung-Wonju National University, Gangneung, Republic of Korea
\item \Idef{org41}Gauhati University, Department of Physics, Guwahati, India
\item \Idef{org42}Helmholtz-Institut f\"{u}r Strahlen- und Kernphysik, Rheinische Friedrich-Wilhelms-Universit\"{a}t Bonn, Bonn, Germany
\item \Idef{org43}Helsinki Institute of Physics (HIP), Helsinki, Finland
\item \Idef{org44}High Energy Physics Group,  Universidad Aut\'{o}noma de Puebla, Puebla, Mexico
\item \Idef{org45}Hiroshima University, Hiroshima, Japan
\item \Idef{org46}Hochschule Worms, Zentrum  f\"{u}r Technologietransfer und Telekommunikation (ZTT), Worms, Germany
\item \Idef{org47}Horia Hulubei National Institute of Physics and Nuclear Engineering, Bucharest, Romania
\item \Idef{org48}Indian Institute of Technology Bombay (IIT), Mumbai, India
\item \Idef{org49}Indian Institute of Technology Indore, Indore, India
\item \Idef{org50}Indonesian Institute of Sciences, Jakarta, Indonesia
\item \Idef{org51}INFN, Laboratori Nazionali di Frascati, Frascati, Italy
\item \Idef{org52}INFN, Sezione di Bari, Bari, Italy
\item \Idef{org53}INFN, Sezione di Bologna, Bologna, Italy
\item \Idef{org54}INFN, Sezione di Cagliari, Cagliari, Italy
\item \Idef{org55}INFN, Sezione di Catania, Catania, Italy
\item \Idef{org56}INFN, Sezione di Padova, Padova, Italy
\item \Idef{org57}INFN, Sezione di Roma, Rome, Italy
\item \Idef{org58}INFN, Sezione di Torino, Turin, Italy
\item \Idef{org59}INFN, Sezione di Trieste, Trieste, Italy
\item \Idef{org60}Inha University, Incheon, Republic of Korea
\item \Idef{org61}Institut de Physique Nucl\'{e}aire d'Orsay (IPNO), Institut National de Physique Nucl\'{e}aire et de Physique des Particules (IN2P3/CNRS), Universit\'{e} de Paris-Sud, Universit\'{e} Paris-Saclay, Orsay, France
\item \Idef{org62}Institute for Nuclear Research, Academy of Sciences, Moscow, Russia
\item \Idef{org63}Institute for Subatomic Physics, Utrecht University/Nikhef, Utrecht, Netherlands
\item \Idef{org64}Institute for Theoretical and Experimental Physics, Moscow, Russia
\item \Idef{org65}Institute of Experimental Physics, Slovak Academy of Sciences, Ko\v{s}ice, Slovakia
\item \Idef{org66}Institute of Physics, Homi Bhabha National Institute, Bhubaneswar, India
\item \Idef{org67}Institute of Physics of the Czech Academy of Sciences, Prague, Czech Republic
\item \Idef{org68}Institute of Space Science (ISS), Bucharest, Romania
\item \Idef{org69}Institut f\"{u}r Kernphysik, Johann Wolfgang Goethe-Universit\"{a}t Frankfurt, Frankfurt, Germany
\item \Idef{org70}Instituto de Ciencias Nucleares, Universidad Nacional Aut\'{o}noma de M\'{e}xico, Mexico City, Mexico
\item \Idef{org71}Instituto de F\'{i}sica, Universidade Federal do Rio Grande do Sul (UFRGS), Porto Alegre, Brazil
\item \Idef{org72}Instituto de F\'{\i}sica, Universidad Nacional Aut\'{o}noma de M\'{e}xico, Mexico City, Mexico
\item \Idef{org73}iThemba LABS, National Research Foundation, Somerset West, South Africa
\item \Idef{org74}Johann-Wolfgang-Goethe Universit\"{a}t Frankfurt Institut f\"{u}r Informatik, Fachbereich Informatik und Mathematik, Frankfurt, Germany
\item \Idef{org75}Joint Institute for Nuclear Research (JINR), Dubna, Russia
\item \Idef{org76}Korea Institute of Science and Technology Information, Daejeon, Republic of Korea
\item \Idef{org77}KTO Karatay University, Konya, Turkey
\item \Idef{org78}Laboratoire de Physique Subatomique et de Cosmologie, Universit\'{e} Grenoble-Alpes, CNRS-IN2P3, Grenoble, France
\item \Idef{org79}Lawrence Berkeley National Laboratory, Berkeley, California, United States
\item \Idef{org80}Lund University Department of Physics, Division of Particle Physics, Lund, Sweden
\item \Idef{org81}Nagasaki Institute of Applied Science, Nagasaki, Japan
\item \Idef{org82}Nara Women{'}s University (NWU), Nara, Japan
\item \Idef{org83}National and Kapodistrian University of Athens, School of Science, Department of Physics , Athens, Greece
\item \Idef{org84}National Centre for Nuclear Research, Warsaw, Poland
\item \Idef{org85}National Institute of Science Education and Research, Homi Bhabha National Institute, Jatni, India
\item \Idef{org86}National Nuclear Research Center, Baku, Azerbaijan
\item \Idef{org87}National Research Centre Kurchatov Institute, Moscow, Russia
\item \Idef{org88}Niels Bohr Institute, University of Copenhagen, Copenhagen, Denmark
\item \Idef{org89}Nikhef, National institute for subatomic physics, Amsterdam, Netherlands
\item \Idef{org90}NRC Kurchatov Institute IHEP, Protvino, Russia
\item \Idef{org91}NRNU Moscow Engineering Physics Institute, Moscow, Russia
\item \Idef{org92}Nuclear Physics Group, STFC Daresbury Laboratory, Daresbury, United Kingdom
\item \Idef{org93}Nuclear Physics Institute of the Czech Academy of Sciences, \v{R}e\v{z} u Prahy, Czech Republic
\item \Idef{org94}Oak Ridge National Laboratory, Oak Ridge, Tennessee, United States
\item \Idef{org95}Ohio State University, Columbus, Ohio, United States
\item \Idef{org96}Petersburg Nuclear Physics Institute, Gatchina, Russia
\item \Idef{org97}Physics department, Faculty of science, University of Zagreb, Zagreb, Croatia
\item \Idef{org98}Physics Department, Panjab University, Chandigarh, India
\item \Idef{org99}Physics Department, University of Jammu, Jammu, India
\item \Idef{org100}Physics Department, University of Rajasthan, Jaipur, India
\item \Idef{org101}Physikalisches Institut, Eberhard-Karls-Universit\"{a}t T\"{u}bingen, T\"{u}bingen, Germany
\item \Idef{org102}Physikalisches Institut, Ruprecht-Karls-Universit\"{a}t Heidelberg, Heidelberg, Germany
\item \Idef{org103}Physik Department, Technische Universit\"{a}t M\"{u}nchen, Munich, Germany
\item \Idef{org104}Politecnico di Bari, Bari, Italy
\item \Idef{org105}Research Division and ExtreMe Matter Institute EMMI, GSI Helmholtzzentrum f\"ur Schwerionenforschung GmbH, Darmstadt, Germany
\item \Idef{org106}Rudjer Bo\v{s}kovi\'{c} Institute, Zagreb, Croatia
\item \Idef{org107}Russian Federal Nuclear Center (VNIIEF), Sarov, Russia
\item \Idef{org108}Saha Institute of Nuclear Physics, Homi Bhabha National Institute, Kolkata, India
\item \Idef{org109}School of Physics and Astronomy, University of Birmingham, Birmingham, United Kingdom
\item \Idef{org110}Secci\'{o}n F\'{\i}sica, Departamento de Ciencias, Pontificia Universidad Cat\'{o}lica del Per\'{u}, Lima, Peru
\item \Idef{org111}Shanghai Institute of Applied Physics, Shanghai, China
\item \Idef{org112}St. Petersburg State University, St. Petersburg, Russia
\item \Idef{org113}Stefan Meyer Institut f\"{u}r Subatomare Physik (SMI), Vienna, Austria
\item \Idef{org114}SUBATECH, IMT Atlantique, Universit\'{e} de Nantes, CNRS-IN2P3, Nantes, France
\item \Idef{org115}Suranaree University of Technology, Nakhon Ratchasima, Thailand
\item \Idef{org116}Technical University of Ko\v{s}ice, Ko\v{s}ice, Slovakia
\item \Idef{org117}Technische Universit\"{a}t M\"{u}nchen, Excellence Cluster 'Universe', Munich, Germany
\item \Idef{org118}The Henryk Niewodniczanski Institute of Nuclear Physics, Polish Academy of Sciences, Cracow, Poland
\item \Idef{org119}The University of Texas at Austin, Austin, Texas, United States
\item \Idef{org120}Universidad Aut\'{o}noma de Sinaloa, Culiac\'{a}n, Mexico
\item \Idef{org121}Universidade de S\~{a}o Paulo (USP), S\~{a}o Paulo, Brazil
\item \Idef{org122}Universidade Estadual de Campinas (UNICAMP), Campinas, Brazil
\item \Idef{org123}Universidade Federal do ABC, Santo Andre, Brazil
\item \Idef{org124}University College of Southeast Norway, Tonsberg, Norway
\item \Idef{org125}University of Cape Town, Cape Town, South Africa
\item \Idef{org126}University of Houston, Houston, Texas, United States
\item \Idef{org127}University of Jyv\"{a}skyl\"{a}, Jyv\"{a}skyl\"{a}, Finland
\item \Idef{org128}University of Liverpool, Liverpool, United Kingdom
\item \Idef{org129}University of Science and Techonology of China, Hefei, China
\item \Idef{org130}University of Tennessee, Knoxville, Tennessee, United States
\item \Idef{org131}University of the Witwatersrand, Johannesburg, South Africa
\item \Idef{org132}University of Tokyo, Tokyo, Japan
\item \Idef{org133}University of Tsukuba, Tsukuba, Japan
\item \Idef{org134}Universit\'{e} Clermont Auvergne, CNRS/IN2P3, LPC, Clermont-Ferrand, France
\item \Idef{org135}Universit\'{e} de Lyon, Universit\'{e} Lyon 1, CNRS/IN2P3, IPN-Lyon, Villeurbanne, Lyon, France
\item \Idef{org136}Universit\'{e} de Strasbourg, CNRS, IPHC UMR 7178, F-67000 Strasbourg, France, Strasbourg, France
\item \Idef{org137}Universit\'{e} Paris-Saclay Centre d'Etudes de Saclay (CEA), IRFU, D\'{e}partment de Physique Nucl\'{e}aire (DPhN), Saclay, France
\item \Idef{org138}Universit\`{a} degli Studi di Foggia, Foggia, Italy
\item \Idef{org139}Universit\`{a} degli Studi di Pavia, Pavia, Italy
\item \Idef{org140}Universit\`{a} di Brescia, Brescia, Italy
\item \Idef{org141}Variable Energy Cyclotron Centre, Homi Bhabha National Institute, Kolkata, India
\item \Idef{org142}Warsaw University of Technology, Warsaw, Poland
\item \Idef{org143}Wayne State University, Detroit, Michigan, United States
\item \Idef{org144}Westf\"{a}lische Wilhelms-Universit\"{a}t M\"{u}nster, Institut f\"{u}r Kernphysik, M\"{u}nster, Germany
\item \Idef{org145}Wigner Research Centre for Physics, Hungarian Academy of Sciences, Budapest, Hungary
\item \Idef{org146}Yale University, New Haven, Connecticut, United States
\item \Idef{org147}Yonsei University, Seoul, Republic of Korea
\end{Authlist}
\endgroup